\documentclass{article}

\usepackage{arxiv}

\usepackage[utf8]{inputenc} 
\usepackage[T1]{fontenc}    
\usepackage{hyperref}       
\usepackage{url}            
\usepackage{booktabs}       
\usepackage{amsfonts}       
\usepackage{nicefrac}       
\usepackage{microtype}      
\usepackage{lipsum}
\usepackage{graphicx}
\usepackage[T1]{fontenc} 
\usepackage{graphicx,color}
\usepackage[dvipsnames]{xcolor}
\usepackage{soul}
\usepackage{url}
\usepackage{subcaption}
\usepackage{amssymb}
\usepackage{float}
\usepackage{pdf14}
\usepackage{tikz}
\usepackage{placeins}

\newcommand{\Like}{\mathcal{L}}
\usepackage{color}

\newcommand{\changes}[1]{\textcolor{black}{#1}}
\newcommand{\minorchanges}[1]{\textcolor{black}{#1}}
\newcommand{\editorchanges}[1]{\textcolor{black}{#1}}
\newcommand{\secondeditorchanges}[1]{\textcolor{black}{#1}}
\graphicspath{ {./img/} }

\title{Neural network reconstructions for the Hubble parameter, growth rate and distance modulus}

\author{
  Isidro G\'omez-Vargas \\
  ICF, Universidad Nacional Aut\'onoma de M\'exico,\\
  62210, Cuernavaca, Morelos, M\'exico.\\
  CICATA-Legaria, Instituto Polit\'ecnico Nacional\\
  11500, Ciudad de M\'exico,  M\'exico.\\
  \texttt{igomez@icf.unam.mx} \\
  \And
  Ricardo Medel Esquivel\\
  CICATA-Legaria, Instituto Polit\'ecnico Nacional, \\
   11500, Ciudad de M\'exico,  M\'exico.\\
  ICF, Universidad Nacional Aut\'onoma de M\'exico,\\
  62210, Cuernavaca, Morelos, M\'exico.\\
  \texttt{rmedel@ipn.mx} \\
  \And
  Ricardo Garc\'ia Salcedo$^*$ \\
 CICATA-Legaria, Instituto Polit\'ecnico Nacional \\
  11500, Ciudad de M\'exico,  M\'exico.\\
  $^*$University of Guanajuato on sabbatical stay.\\
  \texttt{rigarcias@ipn.mx} \\
  \AND
  J.~Alberto V\'azquez$^*$ \\
  ICF, Universidad Nacional Aut\'onoma de M\'exico,\\
  62210, Cuernavaca, Morelos, M\'exico.\\
  \texttt{javazquez@icf.unam.mx} \\}

\begin{document}
\maketitle
\begin{abstract}
This paper introduces a \minorchanges{new approach} to reconstruct cosmological functions using artificial neural networks based on observational measurements with minimal theoretical and statistical assumptions. By using neural networks, we can generate computational models of observational datasets, and then we compare them with the original ones to verify \minorchanges{the consistency of our method}. This methodology is applicable to even small-size datasets. In particular, we \minorchanges{test the proposed method with} data coming from cosmic chronometers, $f\sigma_8$ measurements, and the distance modulus of the Type Ia supernovae. Furthermore, we introduce a first approach to generate synthetic covariance matrices through a variational autoencoder, using the systematic covariance matrix of the Type Ia supernova compilation.


\end{abstract}


\section{Introduction}
\label{sec:intro}
    
    One of the biggest challenges for the cosmological community is the explanation of the current accelerated Universe expansion. A theoretical conception, commonly called Dark Energy (DE), is introduced to explain this mysterious phenomenon and whose nature is still unraveled \cite{copeland2006dynamics, amendola2010dark, ruiz2010dark}.
    The standard model of cosmology, or simply $\Lambda$CDM, is the homogeneous and isotropic cosmological model whose material content is as follows: ordinary matter, the simplest form of Dark Energy known as cosmological constant $\Lambda$ and, finally, a key component for the formation of structures in the Universe called Cold Dark Matter (CDM). It has had significant achievements, such as being in excellent agreement with most of the currently available data, for example, measurements coming from the Cosmic Microwave Background radiation \cite{aghanim2020planck}, Supernovae Ia (SNeIa) \cite{betoule2014improved}, Cosmic Chronometers (CC) \cite{stern2010cosmic} and Baryon Acoustic Oscillations (BAO) \cite{alam2017clustering}. Nevertheless, the $\Lambda$CDM model has its drawbacks: on theoretical grounds, the cosmological constant faces several problems, i.e., fine-tuning and cosmic coincidence \cite{sahni2002cosmological, peebles2003cosmological}, and from an observational point of view, it also suffers to the so-called  Hubble tension, a measurement disagreement of the Hubble parameter $H_0$ among different datasets \cite{feeney2018clarifying}. The presence of these issues opens the possibility to extensions beyond the standard cosmological model by considering, for instance, a dynamical DE, modifications to the general theory of relativity \cite{joyce2016dark} or other approaches.
    
    The search for possible signatures for cosmological models beyond the $\Lambda$CDM has led to the creation of an impressive set of high accuracy surveys, already underway or being planned \cite{tyson2002large, aghamousa2016desi, amendola2018cosmology}, to gather a considerable amount of information that constrains the properties of the universe. A viable cosmological model that leads to the current accelerating universal expansion is demanded to comply with all the relevant observational data. Extensions to the cosmological constant that allow a redshift-dependent equation-of-state (EoS) $w(z)$ include extra dimensions \cite{brax2003cosmology}, modified gravity \cite{clifton2012modified}, scalar fields \cite{vazquez2020bayesian}, scalar-tensor theories with non-minimal derivative coupling to the Einstein tensor \cite{quiros2018phantom}, graduated dark energy \cite{Akarsu:2019hmw}, just to mention a few. 
    However, in the absence of a fundamental and definitive theory of dark energy, a time-dependent behavior can also be investigated by choosing an EoS mathematically appealing or a parameterized form in a simple way; examples of these forms in terms of redshift include a Taylor expansion \cite{chevallier2001accelerating}, polynomial \cite{sendra2012supernova}, logarithmic \cite{odintsov2018cosmological}, oscillatory \cite{Tamayo:2019gqj, liu2009testing}, a combined form of them \cite{Arciniega:2021ffa} or in terms of cosmic time \cite{Akarsu:2015yea}.
    Nonetheless, the \textit{a priori} assumption of a specific theoretical model may lead to misleading model-dependent results regardless of the dark energy properties. Hence, instead of committing to a particular cosmological model, the non-parametric inference techniques allow extract information directly from the data to detect features within cosmological functions, for instance, $w(z)$. The main aim of a non-parametric approach is to infer (reconstruct) an unknown quantity based mainly on the data and make as few assumptions as possible \cite{sahni2002cosmological, wasserman2006all}.
    Several non-parametric techniques are used to perform model-independent reconstructions for cosmological functions directly from the data, such as histogram density estimators \cite{sahni2006reconstructing}, Principal Component Analysis (PCA) \cite{sharma2020reconstruction}, smoothed step functions \cite{gerardi2019reconstruction}, gaussian processes \cite{williams2006gaussian, Keeley:2020aym, l2020defying, mukherjee2020revisiting}, Simulation Extrapolation method (SIMEX) \cite{montiel2014nonparametric} and Bayesian nodal free-form methods \cite{Vazquez:2012ce, hee2017constraining}.
    
    After the reconstruction is performed, the function can be considered as a new model to look for possible deviations from the standard $\Lambda$CDM. 
    In other words, the result of a \changes{model-independent} reconstruction may be used to analyze its similarity with different theoretical models and, therefore, to select its best description for the data.
    There are several examples of \changes{model-independent} reconstructions of cosmological functions, some of them focus on dark energy features \cite{sahni2006reconstructing, holsclaw2010nonparametric, zhao2017dynamical, gerardi2019reconstruction}, on the cosmic expansion \cite{montiel2014nonparametric}, deceleration parameter \cite{mukherjee2020revisiting},  growth rate of structure formation \cite{l2020defying}, luminosity distance \cite{wei2017improved, lin2019non} and primordial power spectrum \cite{Vazquez:2012ux, Handley:2019fll}, among many others. 
    
    On the other hand, the recent increase in computing power and the vast amount of coming data have allowed the incursion of machine learning methods as analysis tools in observational cosmology \cite{lin2017does, peel2019distinguishing, arjona2020can, wang2020machine, gomez2021neural, Chacon:2021sil}. In this work, we focus on the computational models called Artificial Neural Networks (ANNs). They have been used in a variety of applications, such as image analysis \cite{dieleman2015rotation, ntampaka2019deep}, N-body simulations \cite{rodriguez2018fast, he2019learning} and statistical methods \cite{auld2007fast, alsing2019fast, li2019model, hortua2020constraining, hortua2020parameter}.
    
    In a similar way that \changes{model-independent} reconstructions are used to recover the baseline function, the main goal of this paper is to \minorchanges{propose a new method based on} artificial neural networks using solely the current datasets with the most minimal theoretical assumptions. Here, we refer to the neural networks output as \changes{model-independent} reconstructions because they do not incorporate any \textit{a priori} cosmological assumption to generate the model from the datasets.
    This work is similar to previous research in which neural networks produce reconstructions of cosmological functions \cite{escamilla2020deep, wang2020reconstructing, dialektopoulos2021neural}. However, the novel differences here are the exploration of more cosmological datasets, the null consideration of a fiducial cosmology in the reconstructions, the exclusive use of the observational data to train the neural networks (even if they are small), and the new treatment to the non-diagonal error covariance matrices.

    A benefit of using well-trained neural networks is that these do not consider a fiducial cosmology; the data generated can be assumed as new observations of the exact nature of the original dataset. Another advantage of neural networks over other standard interpolation techniques is that, due to their nonlinear modeling capabilities, these do not require consideration of any statistical distribution for the data. In addition, neural networks also allow us to generate computational models for the errors of the observational datasets; when the errors have no correlations (diagonal covariance matrices), we develop a single neural network model that considers measurements and errors from the original dataset. However, we must generate a different neural model when the error matrices are non-diagonal. We show that our methodology can apply to several astronomical datasets, including full covariance matrices with correlations among measurements, for which we introduce a special treatment with variational autoencoders. 

    The rest of the paper has the following structure. In Section \ref{sec:background}, we  briefly introduce the cosmological and statistical concepts used throughout this work: cosmological models, functions and observations in Section \ref{sec:cosmo}; a short summary of Bayesian inference in Section \ref{sec:bayes} and an overview of neural networks in section \ref{sec:ann}. Section \ref{sec:methodology} describes the methodology used during the neural network training to generate computational models based on cosmological data. Section \ref{sec:results} contains our results, in Section \ref{sec:reconstructions} we show the generation of \changes{model-independent reconstructions} using neural networks from observational measurements of the Hubble distance $H(z)$, a combination of the growth rate of cosmological perturbations times the matter power spectrum normalization $f\sigma_8(z)$ and the distance modulus $\mu(z)$ along with its covariance matrix. In Section \ref{sec:results2}, we use \changes{Bayesian inference on two cosmological models to check the consistency of our reconstructions in comparison with the original data and the expected values of the cosmological parameters}. Finally, in Section \ref{sec:conclusions} we expose our final comments. Furthermore, within the appendices, \changes{a brief description of feedforward neural networks and variational autoencoders is included, as well as the training process used for the networks and our experimental method to learn the covariance matrix.}

\section{Cosmological and statistical background}
\label{sec:background}
   
   This section introduces the cosmological models, functions, and datasets used throughout this work. \minorchanges{The datasets are used to develop the model-independent reconstructions with our method and the cosmological models are used to compare these reconstructions with the theoretical predictions.} We also provide a brief overview of the relevant concepts of Bayesian inference\minorchanges{, which we use as a consistency test for the results of our neural network reconstructions,} and of the essential elements of Artificial Neural Networks\minorchanges{, which are the core of our proposed method}. Throughout this paper we use the geometric unit system where $ \hbar = c = 8\pi G = 1$.

\subsection{Cosmological models and datasets} 
    \label{sec:cosmo}
    \subsection*{Models}
    
    The Friedmann equation describing the late-time dynamical evolution for a flat-$\Lambda$CDM model can be written as: 
    \begin{equation}
        H(z)^2 = H_0^2\left [\Omega_{m,0}(1+z)^3 + (1-\Omega_{m,0})\right],
        \label{eq:hzlcdm}
    \end{equation} 
    where $H$ is the Hubble parameter and $\Omega_{m}$ is the matter density parameter, subscript  0  attached  to any quantity denotes its present $(z=  0)$  value. In this case, the DE EoS is $w(z) = -1$.
     
    A step further to the standard model is to consider the dark energy being dynamic, where the evolution of its EoS is usually parameterized. A commonly used form of $w(z)$ is to take into account the next contribution of a Taylor expansion in terms of the scale factor  $w(a)= w_0 + (1-a)w_a$ or in terms of redshift $w(z) = w_0 + \frac{z}{1+z} w_a$; we refer to this model as CPL \cite{chevallier2001accelerating, linder2003exploring}.
    The parameters $w_0$ and $w_a$ are real numbers such that at the present epoch  $w|_{z=0}=w_0$  and $dw/dz|_{z=0}=-w_a$; we recover $\Lambda$CDM when $w_0 = -1$ and $w_a=0$. Hence the Friedmann equation for the CPL parameterization turns out to be:
    \begin{equation}
        H(z)^2 = H_0^2\left[ \Omega_{m,0}(1+z)^3 + (1-\Omega_{m,0})(1+z)^{3(1+w_0+w_a)} e^{-\frac{3w_a z}{1+z}}\right].
        \label{eq:hzcpl}
    \end{equation}
  

    \subsection*{Datasets}
    \textbf{Cosmic chronometers} (CC) are galaxies that evolve slowly and allow direct measurements of the Hubble parameter $H(z)$. These measurements have been collected over several years \cite{jimenez2003constraints, simon2005constraints, stern2010cosmic, moresco2012new, zhang2014four, moresco2015raising, moresco20166, ratsimbazafy2017age}, and now $31$ data points are available within redshifts between $0.09$ and $1.965$, along with their {independent statistical errors}.
    \\
 
    The \textbf{growth rate measurement} is usually
    referred to the product of $f\sigma_8(a)$ where $ f (a) \equiv d \ln \delta(a)/d\ln a $ is the growth rate of cosmological perturbations given by the density contrast $\delta(a)\equiv \delta \rho/\rho$, being $\rho$ the energy density and  
    $\sigma_8$ the normalization of the power spectrum on scales within spheres of $8h^{-1}$Mpc \cite{said2020joint}. 
    Therefore, the observable quantity $f\sigma_8(a)$, or equivalently $f\sigma_8(z)$, is obtained by solving numerically:
    \begin{equation}
        f\sigma_8 (a) =  a\frac{\delta'(a)}{\delta(1)} \sigma_{8,0}.
    \end{equation}
    The  $f{\sigma_8}$ data are obtained through the peculiar velocities from Redshift Space Distortions (RSD) measurements \cite{kaiser1987clustering} observed in redshift survey galaxies or by weak lensing \cite{amendola2008measuring}, where the density perturbations of the galaxies are proportional to the perturbations of matter.
    \secondeditorchanges{We used an extended version of the Gold-2017 compilation available in \cite{sagredo2018internal}, which includes $22$ independent measurements of $f\sigma{_8}(z)$ with their statistical errors obtained from redshift space distortion measurements across various surveys (see references therein); the authors explain that the data used from the $f\sigma{_8}$ combination has been shown to be unbiased.
    Table I of \cite{sagredo2018internal} contains the $f\sigma{_8}(z)$ measurements along with their standard deviations used in this work to form our training dataset. In the same Table, it is indicated the reference matter density parameter $\Omega_{m,0}$ for each measurement and other details of the dataset. }

    \textbf{Supernovae} (SNeIa).
    The SNeIa dataset used in this work corresponds to the Joint Lightcurve Analysis (JLA), a compilation of 740 Type Ia supernovae. It is available in a binned version that consists of 31 data points with a covariance matrix $C_{JLA} \in \mathbb{R}^{31 \times 31}$ related to the systematic measurement errors \cite{betoule2014improved}. As a proof of the concept, we focused on the binned version because, even though the treatment of a matrix in $\mathbb{R}^{740 \times 740}$ from the entire dataset is a straightforward process, it is very computationally expensive (see Appendix \ref{sec:appendix_vae_train} for details). However, it can be implemented on more powerful computers.
    
    Let us assume a spatially flat universe, for which the relationship between  the luminosity distance $d_L$ and  the comoving distance $D(z)$ is given by: 
      \begin{equation}
         d_L (z) = \frac{1}{H_0}(1+z)D(z), \qquad {\rm with }\qquad D(z) = H_0\int \frac{dz}{H(z)}.
         \label{eq:dL}
      \end{equation}

      \secondeditorchanges{Using $d_L$ defined in Equation \ref{eq:dL}, and considering that the distance is expressed in Mega parsecs, the distance modulus is defined as follows:}
       \secondeditorchanges{
       \begin{equation}
           \mu(z) = m - M = 5 \log_{10} d_L(z) + 25,
           \label{eq:mu_th}
       \end{equation}
       }
       \secondeditorchanges{where $m$ is the apparent magnitude and $M$ refers to the absolute magnitude. According to Ref. \cite{betoule2014improved}, in order to use the JLA binned data, and to perform the Bayesian parameter estimation, we need to apply the following likelihood:}
       \secondeditorchanges{
       \begin{equation}
           \log \Like = r^T \cdot C_{\mathrm{JLA}}^{-1} \cdot r,
           \label{eq:loglikeSN}
       \end{equation}
       }
       \secondeditorchanges{
       where $r = \mu_b - M - 5\log_{10} d_L (z)$ and $\mu_b$ is the distance modulus obtained from the binned JLA dataset.}
       
       \secondeditorchanges{
       We can use the definition for the theoretical distance modulus from Equation \ref{eq:mu_th} and obtain $r = \mu_b - \mu(z) + (25 - M)$, and for simplicity, we fixed $M$ because the prior knowledge suggests a constant value \cite{amendola2010dark}. However, reference \cite{betoule2014improved} warns about the importance of treating the absolute magnitude $M$ as a free parameter in the Bayesian inference when using the binned dataset to avoid any potential issues with the estimated value of the Hubble parameter. Nevertheless, as our main aim is to use the Bayesian inference as a proof of the concept for our methodology and not to draw a cosmological conclusion from the results, we have fixed  $M$ to the same value in all the tests for the sake of simplifying computations with the data and their covariance matrix.}

     \noindent 
     \editorchanges{Details about the calibration of the Type IA supernovae binned dataset, and its covariance matrix used in this work are contained in appendices E and F of the Reference \cite{betoule2014improved}}. 
\subsection{Bayesian inference}     
        \label{sec:bayes}
      Given a set of observational data and a mathematical expression for a cosmological model, a conditional probability function can be constructed regarding the model parameters and the observables. There are many ways to infer the combination of parameters that best fit the data. In cosmology, Bayesian inference algorithms have been used prominently \cite{trotta2008bayes, feroz2009, leclercq2018bayesian}; however, methods such as the Laplace approximation \cite{taylor2010analytic}, genetic algorithms \cite{nesseris2012new, arjona2020can}, simulated annealing \cite{hannestad1999stochastic} or particle swarm optimization \cite{prasad2012cosmological} have also been explored. 

      Bayesian statistics is a paradigm in which probabilities are computed given the prior knowledge of the data \cite{padilla2019, medel2021}. It can perform two essential tasks in data analysis: parameter estimation and model comparison. The Bayes' Theorem on which it is based is as follows:
        \begin{equation}
            P(\theta|D)= \frac{P(D|\theta)P(\theta)}{P(D)},
        \end{equation}
      where $D$ represents the observational dataset and $\theta$ is the set of free parameters in the theoretical model. $P(\theta)$ is the prior probability density function and represents the previous knowledge of the parameters. $\Like= P(D|\theta)$ is the likelihood function and indicates the conditional probability of the data $D$ given the parameters $\theta$ of a model. Finally, $P(D)$ is a normalization constant, that is, the likelihood marginalization, and is called the Bayesian evidence. This quantity is very useful in model comparison, for example, it has been used in several papers to compare dark energy models through the Bayes factor and Jeffrey's scale \cite{Vazquez:2012ce, vazquez2020bayesian}.
      
     Considering the datasets described above, we use the following log-likelihoods:
     \begin{equation}
         \log \Like_i = \minorchanges{-\frac{1}{2}}(D^i_{\mathrm{th}}-D^i_{\mathrm{obs}})^T \cdot C_i^{-1} \cdot (D^i_{\mathrm{th}}-D^i_{\mathrm{obs}}), 
     \end{equation}
     where $i=1,2,3$ correspond to the three datasets: cosmic chronometers [$D^{i=1} = H(z)$] and growth rate measurements [$D^{i=2} = f\sigma_8(z)$]. $D_{obs}$ represents the observational measurements, while $D_{\mathrm{th}}$ is the theoretical value for the cosmological models. $C_{i=1}$ and $C_{i=2}$ are diagonal covariance matrices. \secondeditorchanges{The log-likelihood for the SNeIa has been previously defined (see Equation \ref{eq:loglikeSN}).}
        
\subsection{{Artificial neural networks}}    
\label{sec:ann}
    \begin{figure*}
        \captionsetup{justification=raggedright,singlelinecheck=false,font=footnotesize}
        \centering
        \makebox[12cm][c]{
        \includegraphics[trim= 40mm 40mm 40mm 50mm, clip, width=4.8cm, height=4cm]{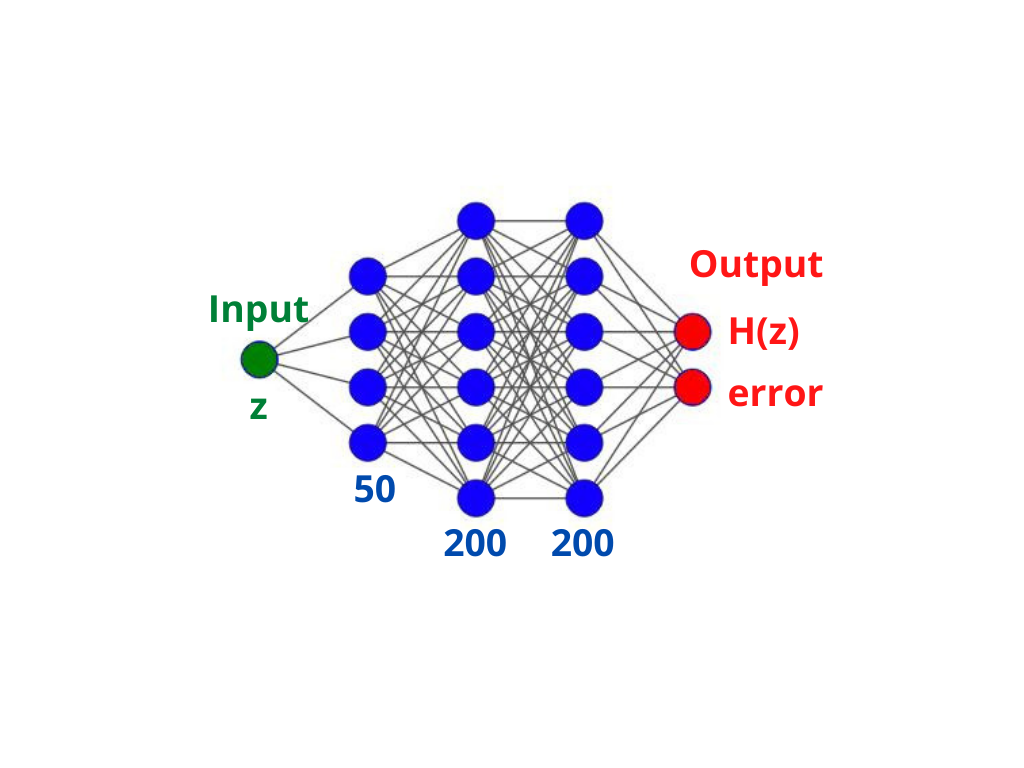}
        \includegraphics[trim= 40mm 40mm 40mm 50mm, clip, width=4.8cm, height=4cm]{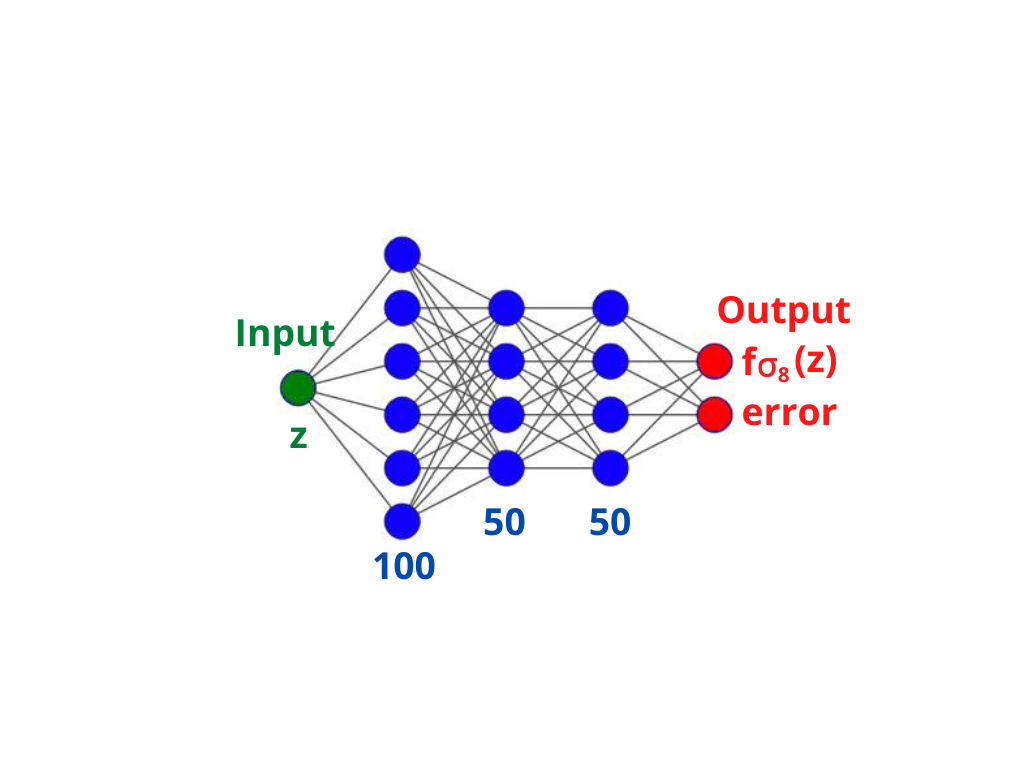}
        \includegraphics[trim= 50mm 40mm 40mm 40mm, clip, width=4.8cm, height=4cm]{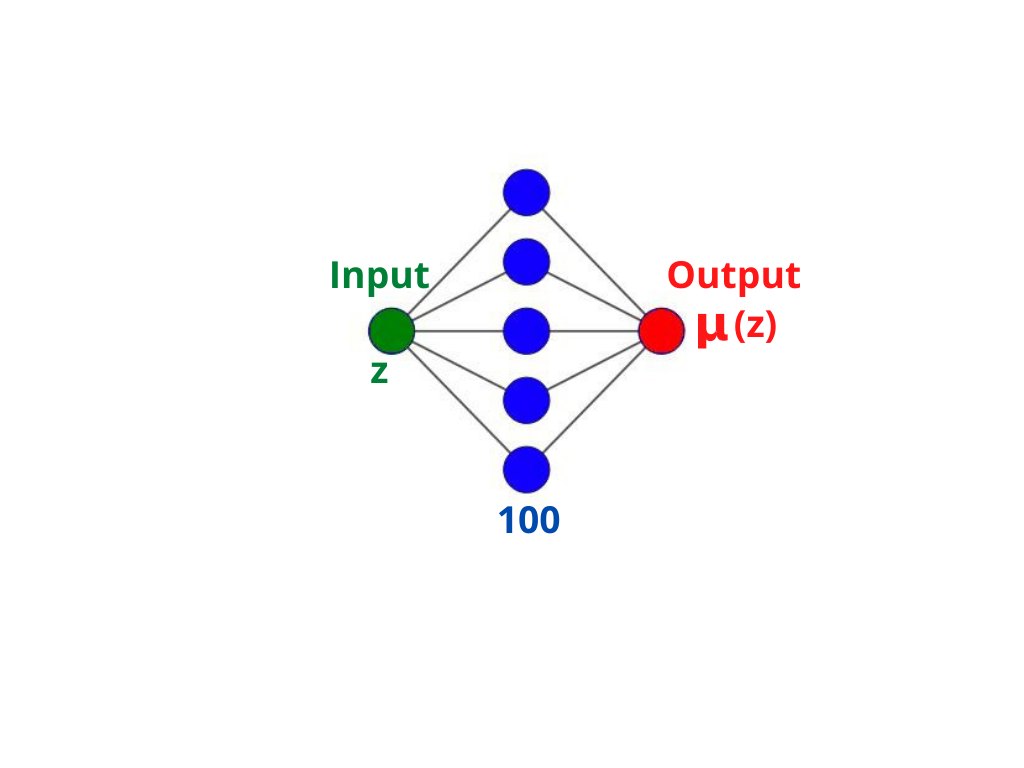}

        }
        \caption{Neural network architectures chosen to model the data from cosmic chronometers (CC), $f{\sigma_8}$ measurements and the JLA SNeIa compilation, respectively; the batch size found for each case was: $16$, $1$ and $1$. In the last architecture, there is only one node in the output layer because the errors are computed with a variational autoencoder (described in the Appendix \ref{sec:appendix_vae_train}) given the original {non-diagonal} covariance matrix of the systematic errors. Blue numbers indicate the nodes in each layer. It is worth mentioning that, in all diagrams, both the redshift functions and the errors in the output layers refer to observational measurements, and no functional form of any cosmology is being considered.}
        \label{fig:arch}
    \end{figure*}
    Artificial Neural Networks (ANNs) are computational models that learn the intrinsic patterns of a dataset. A neural network consists of several sets of neurons or nodes grouped into layers, and between the nodes of different layers some connections are associated with numbers called weights. The training of a neural network aims to find the best values for all the weights to produce a generalization of the data, and this is done through the minimization of an error function (called loss function) that measures the difference between the values predicted by the neural network and the actual values of the dataset (see Appendix \ref{sec:appendix_ann_basics} for more details and in Ref. \cite{juanUniverse} for an introduction to the subject).

    The \textit{Universal Approximation Theorem} \cite{hornik1990universal} states that an Artificial Neural Network with at least one hidden layer with a finite number of neurons can approach any continuous function if the activation function is continuous and nonlinear. Therefore an ANN is capable of learning the intrinsic functions inside cosmological datasets and generating a model based only on the data.
    Two types of artificial neural networks are implemented in this work: Feedforward Neural Networks (FFNN) and AutoEncoders (AE). The FFNN, also called multilayer perceptrons or deep feedforward networks, are the quintessential deep learning models \cite{goodfellow2016deep}. In this type of ANN, the connections and information flow are feed-forward, i.e., from the first to the last layer without loops. These consist of one input layer, at least one hidden layer, and an output layer. The input consists of the dataset's independent variables (or features), while the output contains the dependent variables (or labels). 

    On the other hand, the autoencoders \cite{baldi1989neural} are trained to generate a copy of its input on its output. We use this type of network to learn the errors of a dataset when they conform to a non-diagonal covariance matrix. We use the Variational Autoencoders (VAE). Details about autoencoders are in the Appendix \ref{sec:appendix_autoencoders}.

\section{Methodology}
\label{sec:methodology}
    The datasets in this work contain redshifts, an observable for each redshift and the corresponding statistical errors. Our goal is to generate neural network models for the data despite the complex dependency of these three variables. That is, we take advantage of the ability of neural networks to model the relationship between these variables. Neural networks, with a structure based on multiple neurons and nonlinear activation functions, allow us to generate computational models utterly independent of any existing cosmological model or statistical assumptions.
    
    Even though neural networks are standard in the treatment of large datasets, there is no mathematical constraint in using them for any size of a given dataset, and it is probed in \cite{ingrassia2005neural}; in particular, it is demonstrated that neural models can be used with a total number of weights larger than the number of sample data points. New approaches in neural network research that focus on small datasets are the references \cite{ng2015deep, pasini2015artificial}, and a machine learning field, so-called \textit{few shot learning} \cite{wang2020generalizing}, uses only a few of samples to train the network. It is worth mentioning that using small datasets, although the computing resources are not demanding, it could become challenging to find the hyperparameters that generate an acceptable model. By monitoring the behavior of the loss function both in the training and the validation sets, we can check the equilibrium between the bias and variance to have certainty about the excellent calibration of the neural network.
    
    In all our datasets, we use their lowest and highest redshifts as the limits of the training set, and then we select a random 20\% as the validation set. We do not use a test set due to the small size of the dataset. However, we test several combinations of parameters to select those that generate an excellent neural network model.
    
    For the analysis of cosmic chronometers and $f{\sigma_8} $ measurements, we work with the FFNNs because their diagonal covariance matrices can be arranged into a single column of the same length as the number of observational measures. For these networks, we use the mean squared error (MSE) as a loss function which is a usual selection in regression problems:
    \begin{equation}
        {\rm MSE} = \frac{1}{n} \sum_i^n (Y_i - \hat{Y}_i)^2,
    \end{equation}
    where $Y_i$ is a vector with predictions of the ANN, $\hat{Y}_i$ a vector with the expected values, and $n$ is the number of predictions (or the length of $Y_i$ and $\hat{Y}_i$). 
    In the case of SNeIa data, we use a FFNN to learn the distance modulus and a Variational Autoencoder for the non-diagonal covariance matrix of the systematic errors.
    
    In addition, we implemented the Monte Carlo Dropout (MC-DO) method \cite{gal2015dropout} in all our FFNNs. This method allows the output of a neural network to have an uncertainty associated with it and to generate robust models due to the dropout being a regularization technique. In the last part of Appendix \ref{sec:appendix_ann_basics}, we describe the basic definitions of Dropout and MC-DO.
    
    {We found the best architectures (shown in Figure \ref{fig:arch}) among several combinations of the intrinsic parameters (hyperparameters) of the neural networks. Appendix \ref{sec:appendix_ffnn_train} describes our careful selection of the hyperparameters of the feedforward neural networks, such as epochs, number of nodes, and how we have applied the MC-DO method. On the other hand, Appendix \ref{sec:appendix_vae_train} explains how we configure the VAE neural network, its loss function and other details about its training, with the non-diagonal covariance matrix of the binned JLA compilation.} 

    {Once the neural networks are well trained, \changes{they constitute a model-independent reconstruction, for which we can compare with observations and theoretical predictions. As a consistency test of our neural reconstructions, we perform Bayesian inference for $\Lambda$CDM and CPL models, and the expected posterior probabilities would be very similar between the reconstruction and the original datasets; otherwise, another neural network architecture must be chosen}. We use the following flat priors: for the matter density parameter today  $\Omega_m \in [0.05, \; 0.5]$, for the physical baryon density parameter $\Omega_b h^2 \in [0.02, \; 0.025]$, for the reduced Hubble constant $h \in [0.4 , \; 0.9]$, and for the amplitude of the (linear) power spectrum $\sigma_8 \in [0.6 , \; 1.0]$. When assuming the CPL parameterisation, we use $w_0 \in [-2.0 , \; 0.0]$ and $w_a \in [-2.0 , \; 2.0]$. The $h$ parameter refers to the dimensionless reduced Hubble parameter today ${H}/{100}$kms$^{-1}$Mpc$^{-1}$.}
\section{Results}
\label{sec:results}

 
    From the observational datasets, we have trained the neural networks to reconstruct the Hubble parameter $H(z)$, the growth rate $f{\sigma_8}(z)$ and the distance modulus $\mu(z)$, \changes{their predictions conform the corresponding model-independent reconstructions}. Finally, we have performed the parameter estimation \changes{to test the consistency of the reconstructions}. 
    
    
\subsection{Reconstructions}
\label{sec:reconstructions}
\subsection*{$H(z)$ data}
\begin{figure*}
        \centering
        \captionsetup{justification=raggedright,singlelinecheck=false,font=footnotesize}
        \makebox[11cm][c]{
        \includegraphics[trim= 0mm 0mm 0mm 0mm, clip, width=8.5cm, height=5.5cm]{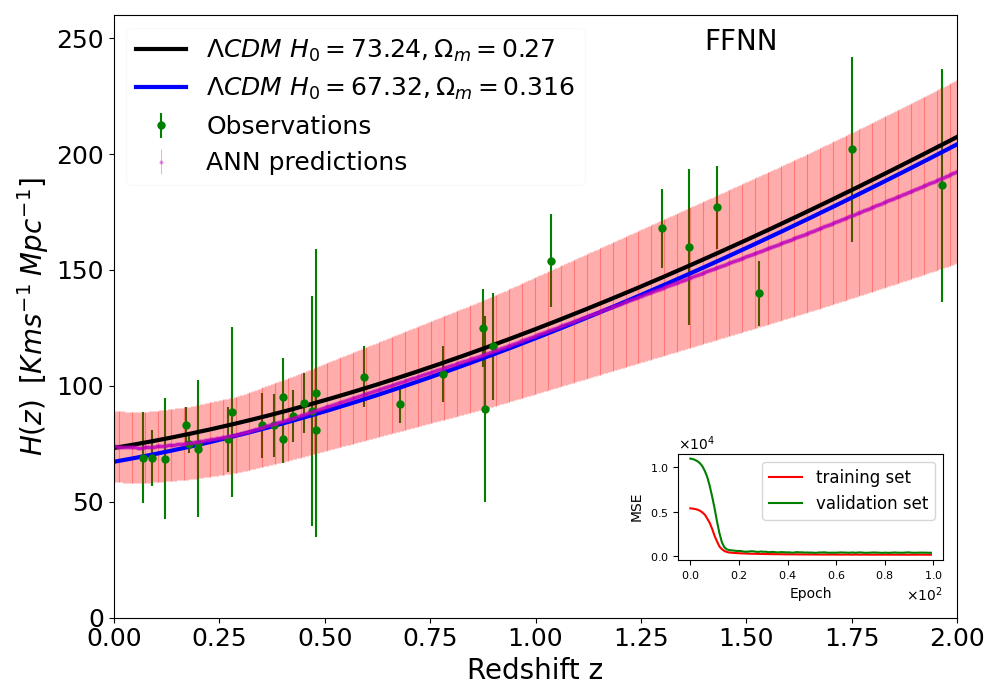}
        \includegraphics[trim= 0mm 0mm 0mm 0mm, clip, width=8.5cm, height=5.5cm]{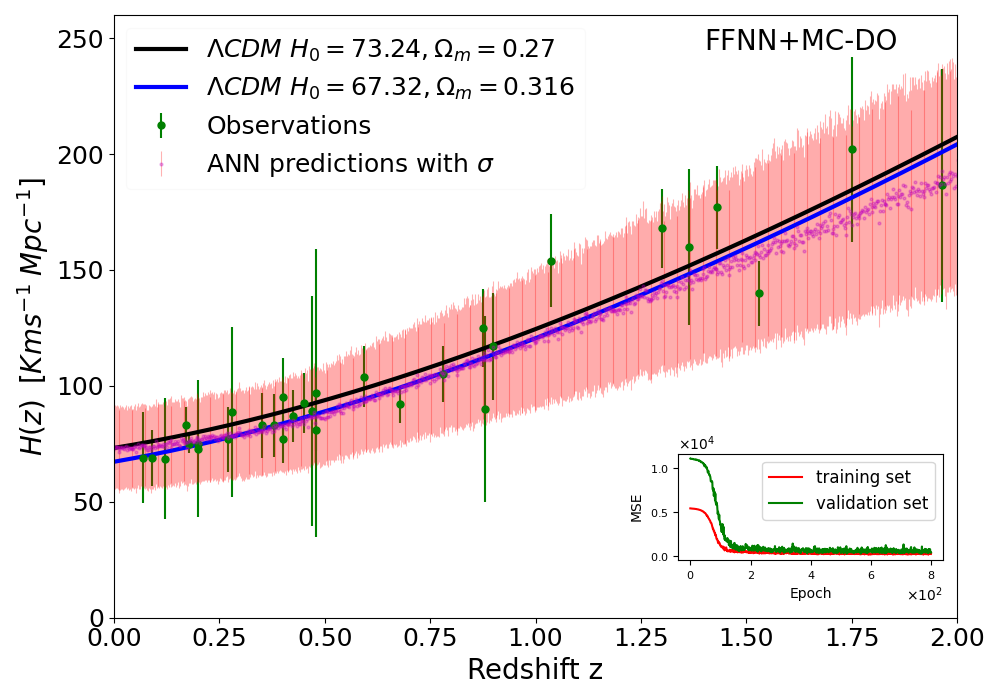}
    }
    \caption{\changes{Hubble distance reconstruction with FFNNs.} \textit{Left}: Purple points represent the FFNN predictions for $H(z)$ along with their error bars in red color. \textit{Right}: Similarly to FFNN but adding MC-DO, we executed $100$ times the Monte Carlo dropout to compute the uncertainties of the predictions. Therefore the purple points are the average predictions of the MC-DO executions, and the red error bars are the uncertainties of the FFNN plus the error predictions (see Equation \ref{eq:sigma}). In both cases, we compare the \changes{neural reconstructions} with the original cosmic chronometers (green bars) and $H(z)$ from $\Lambda$CDM, as shown in the labels. The small panels show the individual behavior of the loss function (MSE) in the training (red) and validation (green) sets, and these plots suggest that it is an excellent neural network model with no overfitting or underfitting.
    }
    \label{fig:recHZ}
    \end{figure*}

     {To visualize the $H(z)$ reconstructions performed by the FFNN using the CC, we generate predictions of $H(z)$ and their corresponding errors given $1000$ different redshifts. In Figure \ref{fig:recHZ} we show the FFNN alone (left) and the FFNN using MC-DO (right), where the original data points with their statistical errors are green, while in magenta the \changes{neural network reconstruction} along with their predicted errors. Also in this figure, we compare the outputs of the neural network models with the theoretical predictions of $\Lambda$CDM using the two values that yield the Hubble tension $H_0 = 73.24 \;$km s$^{-1}$ Mpc$^{-1}$ \changes{and $\Omega_m = 0.27$} coming from the Cepheid variables \cite{riess20162} and, \changes{on the other hand,} $H_0 = 67.40 \;$ km s$^{-1}$ Mpc$^{-1}$ \changes{and $\Omega_m = 0.316$} measured by the Planck mission \cite{aghanim2020planck}.}
    
    \changes{We can notice that the FFNN alone and with MC-DO generate reconstructions in agreement with the theoretical predictions, exclusively based on the observable measurements and their statistical errors. The observational data points are only a few, therefore the scatter of the measurements is underestimated; however, based on their curves for the loss function, we can confirm that the neural networks generate good models. The dispersion in the FFNN+MC-DO reconstruction is higher because it performs statistics on several predictions and includes the uncertainty of the method itself, therefore its results are more robust and reliable than FFNN alone.}
    
    It is worth mentioning that the reconstructions performed by our method with FFNNs are consistent with the $H(z)$ reconstructions of other works performed with Gaussian processes \cite{singirikonda2020model, wang2020machine, mukherjee2021assessment, bonilla2021measurements} and with neural networks \cite{wang2020reconstructing}, where the training dataset consists of $H(z)$ evaluations from a flat $\Lambda$CDM cosmology, redshifts are distributed under a gamma distribution, and errors are produced by an analytical expression \cite{ma2011power}. In this sense, the advantages of our results are that they have no statistical assumptions on the data as Gaussian processes usually do, we do not use either the Friedmann equation or another cosmological equation to augment the datasets, and the neural networks learned directly to model observational errors without imposing some analytical expression beforehand.

\subsection*{$f{\sigma_8}(z)$ data}
\begin{figure*}
        \centering
        \captionsetup{justification=raggedright,singlelinecheck=false,font=footnotesize}
        \makebox[11cm][c]{
        \includegraphics[width=9cm, height=5.2cm]{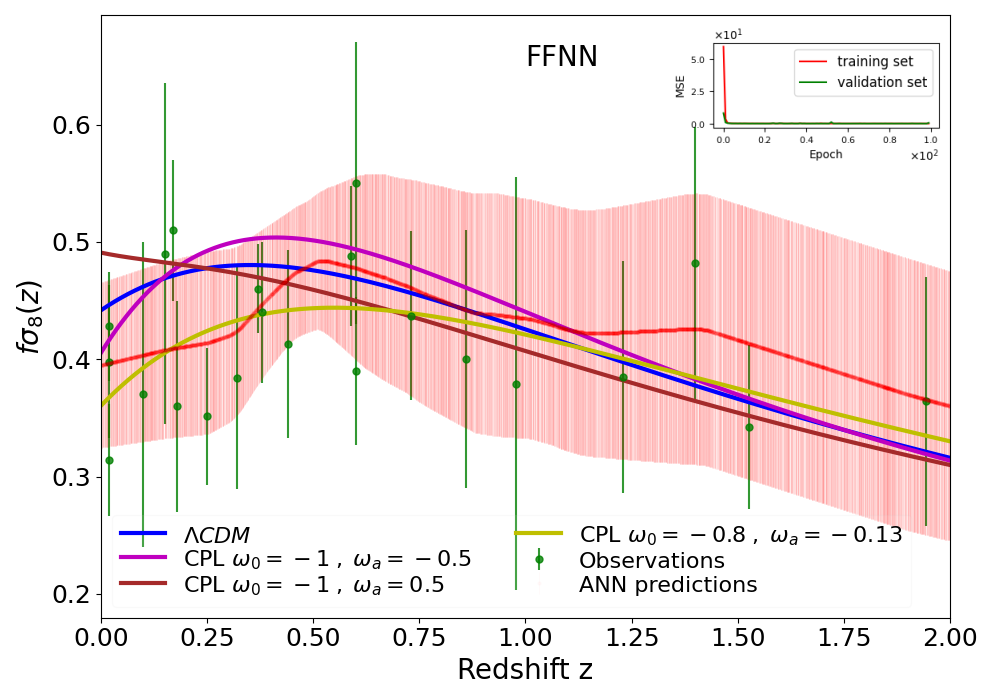}
        \includegraphics[width=9cm, height=5.2cm]{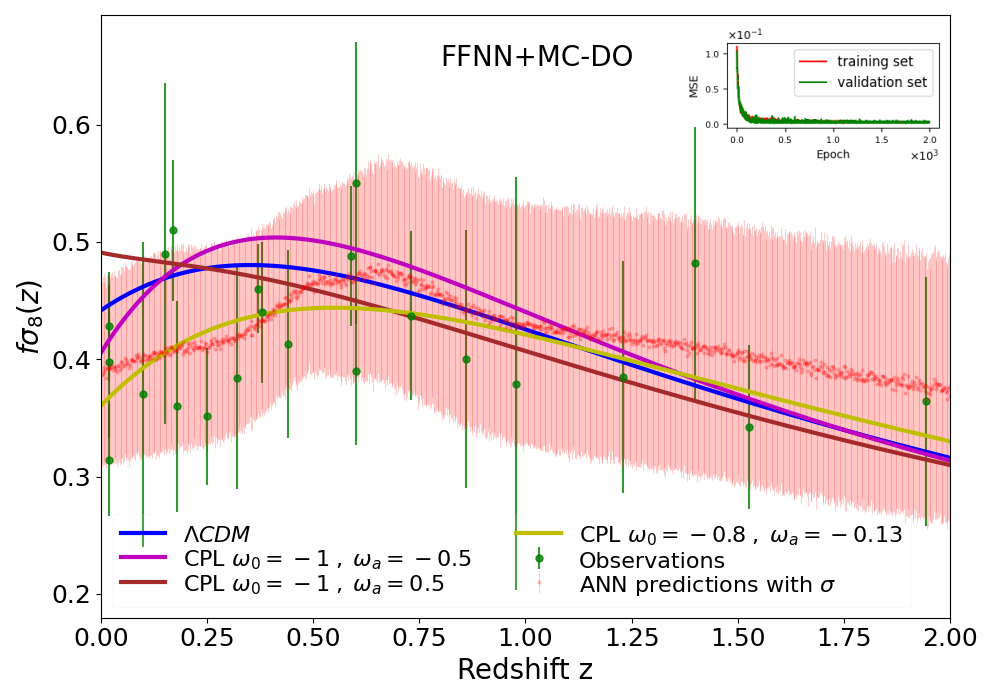} 
        }
        \caption{\changes{Neural reconstructions for} $f\sigma_8(z)$ (red lines) and their respective errors. \changes{The original observations are in green with their \editorchanges{uncertainties}}. \textit{Left}: The $f{\sigma_{8}}$ \changes{reconstruction performed} by FFNN alone. \textit{Right}: $f{\sigma_{8}}$ \changes{reconstruction} predicted by the FFNN using Monte Carlo dropout, the averages of $100$ executions of MC-DO are indicated with the red data and their standard deviations are added to the error predictions. In both cases, the small panels display the behavior of the loss function (MSE) in the training (red curve) and validation (green curve) sets; in this case, these curves also show a good neural network model.}
    \label{fig:recFS8}
    \end{figure*}

    {\editorchanges{We trained} the FFNNs with the extended Gold-2017 compilation of growth rate measurements \editorchanges{and their statistical uncertainties}, we generate $1000$ \changes{predictions from the trained neural nets} to visualize the $f{\sigma_{8}}(z)$ reconstructions. In Figure \ref{fig:recFS8}, we plot the original data with their \editorchanges{uncertainties} (green), while the neural network predictions and their errors are displayed in red (left panel is the FFNN alone and right panel corresponds to FFNN+MC-DO)}. We also draw some curves of $f{\sigma_{8}}(z)$ from the analytical evaluation of the CPL model for different values of $w_0$ and $w_a$.
    We notice that the models are within the reconstructions in both cases. Hence, this dataset by itself may provide loose constraints on the CPL parameters, {mainly because there are very few points and relatively large statistical errors}. However, the values $w_0 = -0.8$ and $w_a = -0.4$ (yellow line) seem to have a better agreement with the reconstruction.
    
    We can analyze Figure \ref{fig:recFS8} to compare the two results.  We can deduce that it is better to use the MC-DO than just the FFNN alone because MC-DO provides a dropout as a regularization technique, avoiding overfitting and producing a more general data model. The small dataset makes for the FFNN alone difficult to learn at redshifts close to zero; however, FFNN+MC-DO performs better in that respect. \changes{Regarding the MC-DO improvement, it can be noticed that in the case of the FFNN method, several data points are outside the reconstruction, while in the reconstruction generated by FFNN+MC-DO only the $f{\sigma_{8}}(z=0.17)=0.51$ point is excluded.}
    {Despite the significant errors and its sparsity, the FFNNs could generate a model consistent with the underlying cosmological theory of the $\Lambda$CDM and CPL models. Moreover, the reconstructions produced by the FFNNs have a similar trend to other \changes{model-independent} reconstructions of $f{\sigma_{8}}(z)$ made by Gaussian processes \cite{l2020defying, yin2019non} with the advantage of letting aside any statistical assumption of the data distribution.}

\subsection*{Distance modulus $\mu (z)$ data}
    %
    {Our reconstruction methodology for the distance modulus differs from those previously presented; in this case, the main aim is modeling the errors of the observational measurements when they are correlated, that is, when the covariance matrix is non-diagonal. For this purpose, we introduce a new method based on a} variational autoencoder (VAE) along with an FFNN to perform the whole neural network modeling for this dataset.
    
\begin{figure}[t]
      \centering
        \includegraphics[width=18cm, height=3.0cm]{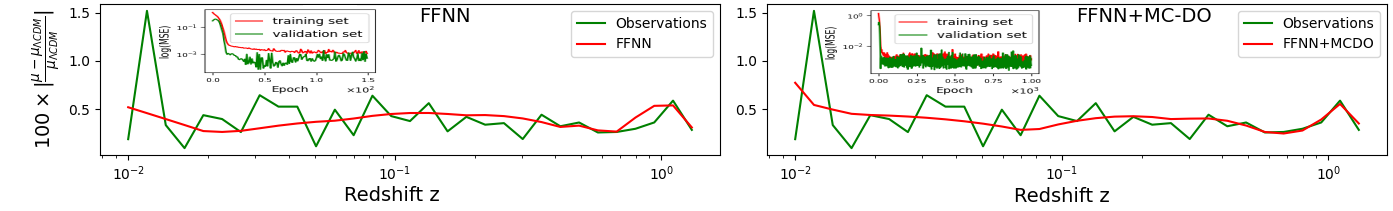}
        %
        \captionsetup{justification=raggedright,singlelinecheck=false,font=footnotesize}
        \caption{\changes{Comparison between the percentage error times 100 between the $\Lambda {\rm CDM}$ theoretical predictions for the distance modulus with the observational measurements and the neural network reconstructions. In the small panels of both figures, the behavior of the loss function, in logarithmic scale, is shown for both the validation (green curve) and training (red curve) sets along with the number of epochs for each case (300 and 1800); it can be seen that after the training process, we obtain acceptable models for the binned JLA dataset. \textit{Left}: With FFNN alone. \textit{Right}: FFNN with Monte Carlo Dropout.} }      
       \label{fig:recJLA}
    \end{figure}

    With the distance modulus \changes{reconstruction, performed by the FFNNs}, we have generated synthetic data points from $31$ log-uniformly distributed redshift values $z \in [0.01, 1.3]$ plus a small Gaussian noise for both the FFNN alone and the FFNN+MC-DO. For comparison, in the Figure \ref{fig:recJLA} \changes{are the percentage differences between the $\Lambda$CDM predictions with the original observations from the binned JLA compilation (in green), and with the neural networks reconstructions (in red).}
    
    We can generate several points at any different values of redshift from the neural network models trained with the distance modulus 
    and model the errors with a VAE neural network (see Appendix \ref{sec:appendix_vae_train} for details of the developed method). \changes{Our motivation for using autoencoders for the covariance matrix is that an autocoder is trained to generate an output of the same nature as the input while encoding a compressed representation in the part between the encoder and the decoder. In addition, if we use a VAE, during training this compressed representation is also sampled through variational inference and, at the end of training, we can know the probability distribution that characterizes it and perform interpolations, sweeping the latent space, to generate new covariance matrices. Furthermore, we can force the dimension of this compressed representation (latent space) to be one-dimensional, for easier interpretation or to map to another 1D distribution.}

\begin{figure}[t]
        \centering
         \captionsetup{justification=raggedright,singlelinecheck=false,font=footnotesize}
        \makebox[11cm][c]{
        \includegraphics[trim= 30mm 0mm 15mm 0mm, clip, width=4.2cm, height=4.1cm]{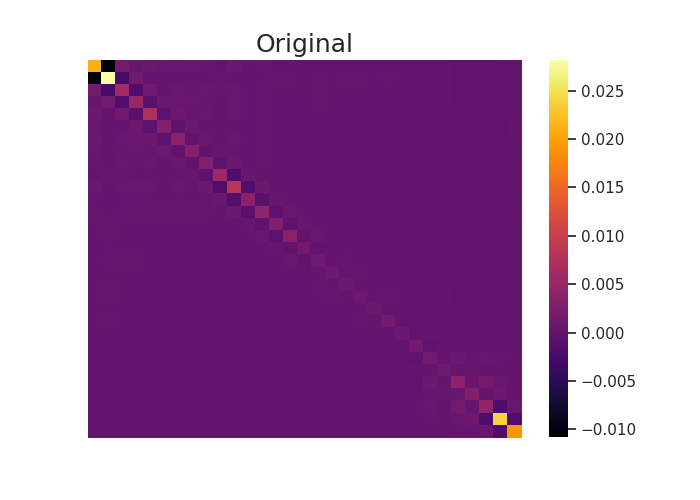}
        \includegraphics[trim=10mm 0mm 10mm 0mm, clip, width=4.5cm, height=4.1cm]{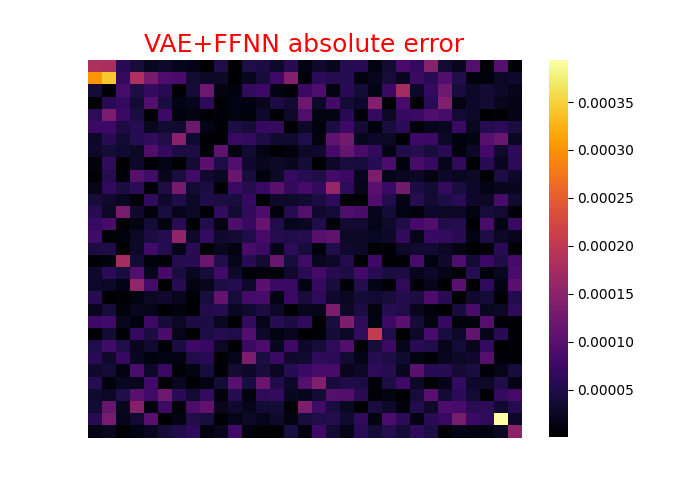}
        \includegraphics[trim=10mm 0mm 10mm 0mm, clip, width=4.5cm, height=4.1cm]{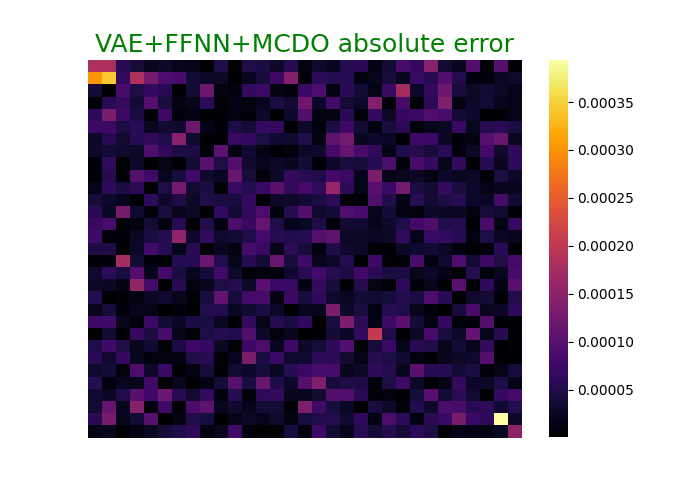}
        }
        \caption{\textit{Left}: Original covariance matrix with systematic errors from JLA compilation (binned version) with 961 entries. \changes{\textit{Middle}: Absolute error for the covariance matrices predicted by the VAE+FFNN concerning the original ones. \textit{Right}: Absolute error for the covariance matrices predicted by the VAE+FFNN with MC-DO.}}
       \label{fig:covmatrix}
    \end{figure}

    A limitation of our method is that the new points, and errors, should correspond to the dimensionality of the matrix, in our case 31.
    Figure \ref{fig:covmatrix} shows the \changes{absolute error for the outputs of the VAE trained with the non-diagonal covariance matrix of the JLA systematic errors; it can be seen that in both cases (VAE+FFNNN and VAE+FFNNN+MC-DO) the differences are two or more orders of magnitude lower than the original matrix; therefore the new matrices are in agreement with the original one. Nonetheless, in Section \ref{sec:results2}, we test these covariance matrices predicted by the neural networks in a Bayesian inference framework to verify whether they are statistically consistent with the original data.}

    From Figure \ref{fig:recJLA}, it can be seen that the \changes{reconstructions} are in better agreement with the $\Lambda$CDM model than to the original data points\changes{; this may occur because when the neural network generates a model for all data points, it underestimates some of the observational variances and focuses more on the similarity of all observations. The FFNN alone has a smaller error in the first prediction, but the FFNN+MC-DO reconstructs the last redshifts better; however, based on the behavior of the loss function, we can say that the computational models generated by the neural networks for the binned JLA compilation are acceptable, both in the case of the FFNN alone, and with MC-DO.}
    
    %
    
\subsection{\changes{Testing reconstructions with Bayesian inference}}
\label{sec:results2}

    \changes{We use a Bayesian inference process for testing the consistency of the reconstructions obtained with the neural networks}. In addition to the three original datasets (cosmic chronometers, $f{\sigma_8}$ measurements, and binned JLA compilation), we have created two datasets for each type of observation from the trained FFNNs with and without MC-DO. As proof of the concept, the new datasets for CC and $f{\sigma_8}$ consist of 50 random uniformly distributed points in redshift. At the same time, for SNeIa, they were 31 log-uniformly distributed in redshift (same size as the original dataset). We also generated its respective covariance matrix for the SNeIa case with the decoder part of the trained VAE.
    We performed the Bayesian inference with the \changes{data from the neural networks reconstructions} and with the original data to \changes{evaluate the quality of the reconstructions}. For this purpose, we analyze the $\Lambda$CDM and \changes{CPL models}. \changes{The idea is that if the neural network reconstructions are satisfactory, the Bayesian estimation of the parameters for the theoretical models should be very similar from those obtained with the original observations, \textit{i.e.}, they should have similar means and standard deviations in the posterior distributions. If this condition is not satisfied, it is necessary to retrain the neural networks or use another hyperparameter configuration.
}
    
\begin{figure*}
    \centering
    \captionsetup{justification=raggedright,singlelinecheck=false,font=footnotesize}
    \makebox[11cm][c]{
            \includegraphics[trim=3mm 0mm 0.0mm 0mm, clip, width=4.2cm, height=4cm]{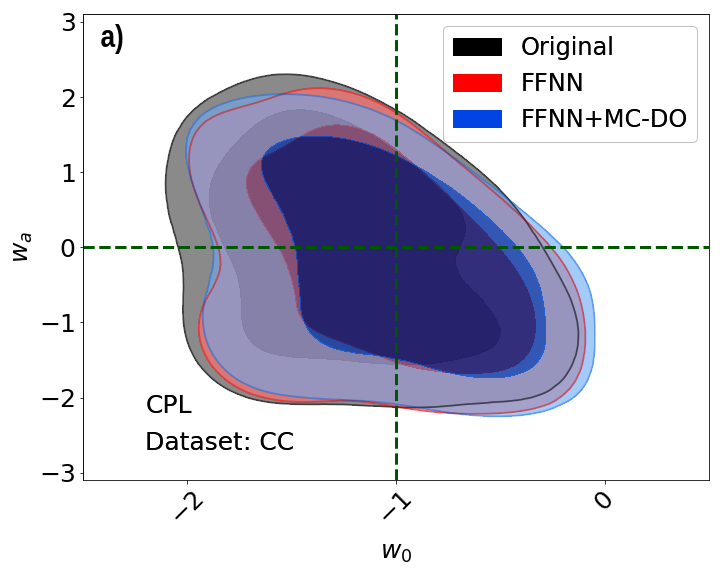}
            \includegraphics[trim=3mm 0mm 0.0mm 0mm, clip, width=4.2cm, height=4cm]{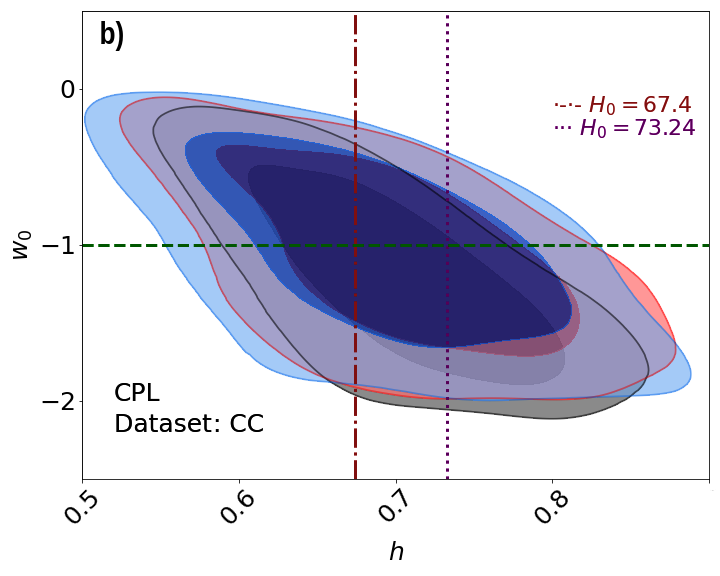}
            \includegraphics[trim=3.5mm 0mm 0.0mm 0mm, clip, width=4.2cm, height=4cm]{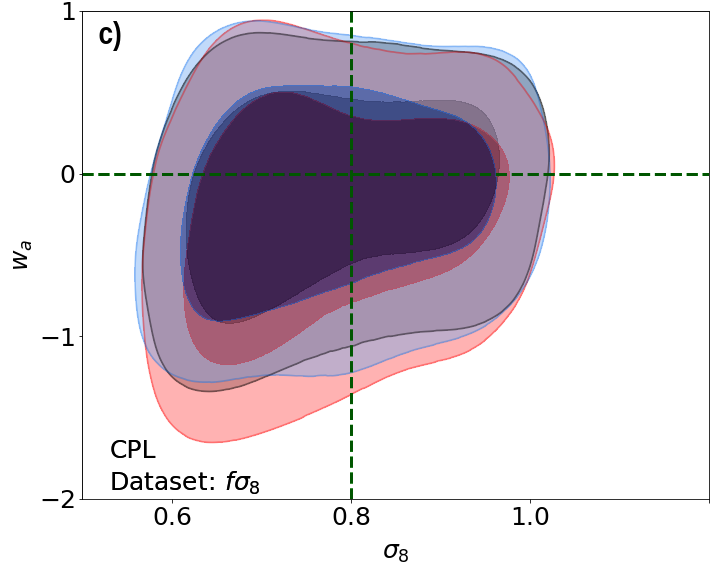}

            }
    \makebox[11cm][c]{
            \includegraphics[trim=3.5mm 0mm 0.0mm 0mm, clip, width=4.2cm, height=4cm]{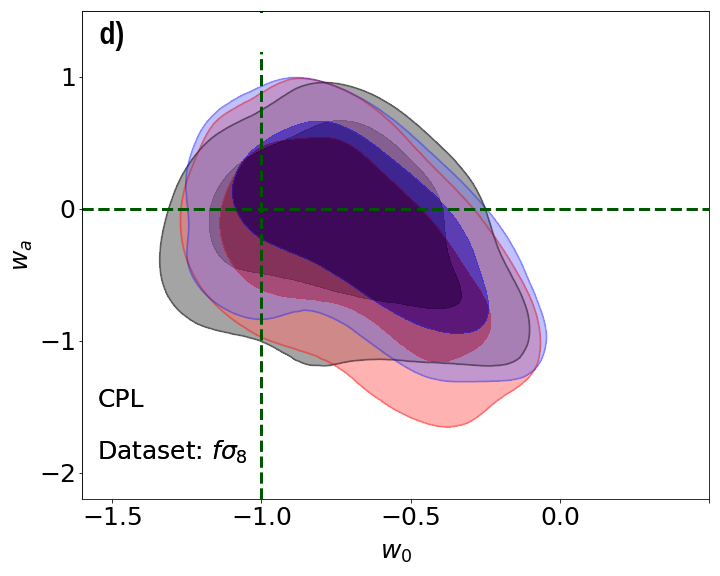}
            \includegraphics[trim=3.5mm 0mm 0.0mm 0mm, clip, width=4.2cm, height=4cm]{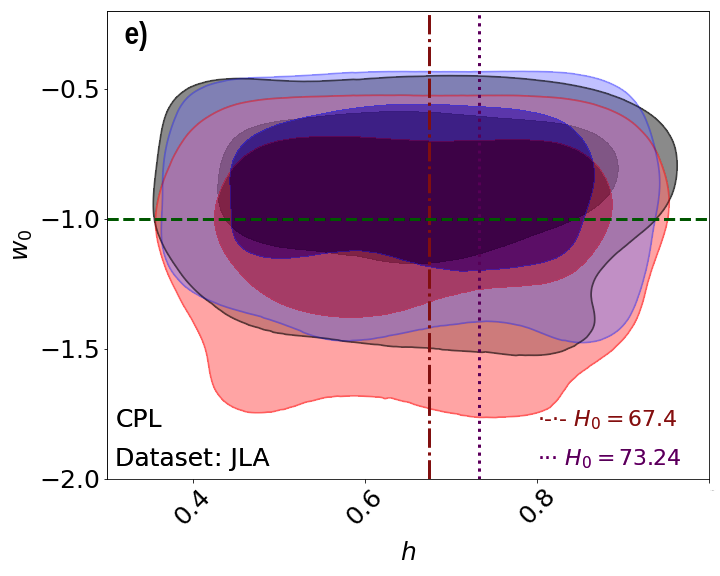}  
             \includegraphics[trim=3.5mm 0mm 0.0mm 0mm, clip, width=4.2cm, height=4cm]{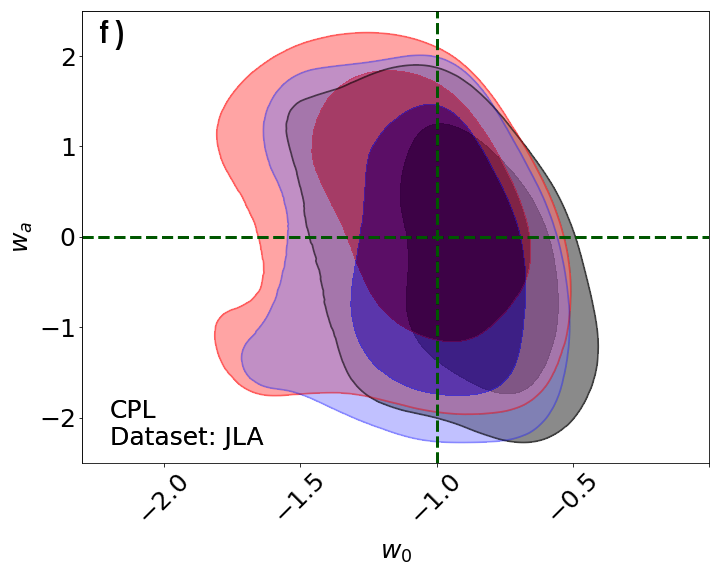}  
            }
   
        \caption{2D marginalized posterior distributions from different combinations of datasets:  original data, \changes{reconstructions data points} from FFNN and FFNN+MC-DO. The green dashed lines  ($w_0=-1$, $w_a=0$) and ($\Omega_1 = 0$, $\Omega_2 = 0$) correspond to the $\Lambda$CDM model. The constraints are plotted with $1\sigma$ and $2\sigma$ confidence contours.}
        \label{fig:posteriorall}
    \end{figure*}

    We have used the data from CC, $f\sigma_8$ measurements, and JLA separately. The most representative results are in Figure \ref{fig:posteriorall}, along with Table \ref{tab:results}, which contains mean values and standard deviations, and they have been sorted according to the datasets used as a source  (original, FFNN, and FFNN+MC-DO), and to the two models involved ($\Lambda$CDM and CPL). Results are displayed for the reduced Hubble parameter $h$, $\sigma_8$, $w_0$ and $w_a$ parameters for the CPL model. In addition, the last column of the Table \ref{tab:results} contains the $-2\ln\Like_{\mathrm{max}}$ of the Bayesian inference process for each case.
    One thing to note is that the neural networks make models that could be thought of as a function $g: z \in \mathbb{R}\rightarrow v\in \mathbb{R}^2$, $v = (f(z), err(f(z)))$, where both $f(z)$ and the error of the observational measurements are being modeled, so when neural network predictions are used to make Bayesian inference, the errors are of the same order of magnitude as the original ones. 
    Before analyzing each scenario separately, it is worth mentioning some generalities in the results. First, it can be noted that when using a single source separately, the constraints are consistent. They all have a similar best fit (maximum likelihood), and secondly, the results agree with the $\Lambda$CDM model.

    \changes{In the case of parameter estimation, displayed in Table \ref{tab:results}, and posterior distributions shown in Figure \ref{fig:posteriorall}, we notice that the best-fit values are mutually contained within their $1\sigma$ standard deviations, in agreement with the $\Lambda$CDM and CPL values. Therefore, the neural network models generated by cosmic chronometers,$f{\sigma_8}(z)$ measurements and distance modulus, through the Bayesian parameter estimation, are statistically consistent with each other.}
    
    %
   

 \begin{table*}
    \scriptsize
    \centering
    \captionsetup{justification=raggedright, singlelinecheck=false, font=footnotesize}
        \begin{tabular}{p{2.0cm} p{1.2cm}p{2.0cm}p{2.2cm}p{2.0cm}p{2.0cm}p{1.4cm}}
    \hline															
    \hline															
    	&		&		\small{\textbf{Datasets:}} 	&	\small{\textbf{CC}}	&		&		&  $\;$	\\
Source	&	Model		&	$h$	&	$w_0$	&	$w_a$	&	$\;$	&	$ -2 \ln \Like_{\mathrm{max}}$	\\
    \hline															
Original	&	 $\Lambda$CDM 	&	$0.678  \pm 0.039$	&	$--$	&	$--$	&	$\;$	&	$14.502$	\\
    &	 CPL 	&	$0.703 \pm 0.064$	&	$-1.223 \pm 0.447$	&	$-0.061 \pm 1.075$	&	$\;$	&	$14.290$ \vspace{0.1cm}	\\

FFNN 	&	$\Lambda$CDM  		&	$0.698 \pm 0.057$	&	$--$	&	$--$	&	$\;$	&	$0.176$	\\    
    &	CPL  		&	$0.703 \pm 0.071$	&	$-1.072 \pm 0.431$	&	$-0.179 \pm 1.025$	&	$\;$	&	$0.042$	\vspace{0.1cm}	\\

FFNN+MC-DO 	&	$\Lambda$CDM  		&	$0.699 \pm 0.063$	&	$--$	&	$--$	&	$\;$	&	$0.346$	\\
    	&	 CPL 	&	$0.689 \pm 0.078$	&	$-1.014 \pm 0.450$	&	$-0.227 \pm 1.003$	&	$\;$	&	$0.284$	 \\
    \hline															
    \hline															
    	&		&	\small{\textbf{Datasets:}} 	&	\small{\textbf{${f\sigma_8}$}}	&		&		&		\\
	&		&		$h$	&	$w_0$	&	$w_a$	&	$\sigma_8$	&		\\
    \hline															
Original 	&	 $\Lambda$CDM 	&	$0.648 \pm 0.147$	&	$--$	&	$--$	&	$0.787 \pm 0.115$	&	$11.932$	\\
    &	 CPL 	&	$0.638 \pm 0.135$	&	$-0.742 \pm 0.264$	&	$-0.144 \pm 0.468$	&	$0.777 \pm 0.111$	&	$11.908$	\vspace{0.1cm}	\\

 FFNN	&	 $\Lambda$CDM 	&	$0.650 \pm 0.144$	&	--	&	$--$	&	$0.694 \pm 0.172$	&	$0.292$	\\
         &	 CPL 	&	$0.648 \pm 0.142$	&	$-0.701 \pm 0.271$	&	$-0.290 \pm 0.540$	&	$0.777 \pm 0.111$	&	$0.284$	\vspace{0.1cm}	\\

 FFNN+MC-DO  	&	  $\Lambda$CDM 	&	$0.651 \pm 0.147$	&	$--$	&	$--$	&	$0.652 \pm 0.170$	&	$0.984$	\\
       	&	 CPL 	&	$0.632 \pm 0.140$	&	$-0.674 \pm 0.270$	&	$-0.156 \pm 0.489$	&	$0.775 \pm 0.110$	&	$0.960$	\\
    \hline															
    \hline															
    	&		&		\small{\textbf{Datasets:}}	&	\small{\textbf{JLA}}	&		&		&	\\
	&		&	$h$	&	$w_0$	&	$w_a$	&	$\;$	&		\\
    \hline	
    
 Original	&	 $\Lambda$CDM 		&	$0.638 \pm 0.146$	&	$--$	&	$--$	&	$\;$	&	$33.214$	\\
      &	CPL  		&	$0.652 \pm 0.141$	&	$-0.901 \pm 0.238$	&	$-0.216 \pm 0.899$	&	$\;$	&	$32.354$	\vspace{0.1cm}	\\
 FFNN	&	 $\Lambda$CDM 		&	$0.645 \pm 0.144$	&	$--$	&	$--$	&	$\;$	&	$14.670$	\\
    &	CPL 	&	$0.640 \pm 0.137$	&	$-1.092 \pm 0.277$	&	$0.287 \pm 0.957$	&	$\;$	&	$13.888$	\vspace{0.1cm}	\\
     
 FFNN+MC-DO	&	 $\Lambda$CDM 	&	$0.643 \pm 0.142$	&	$--$	&	$--$	&	$\;$	&	$16.446$	\\
        &	 CPL 	&	$0.641 \pm 0.135$	&	$-1.037 \pm 0.248$	&	$-0.245 \pm 0.996$	&	$\;$	&	$16.274$ \\
    \hline															
    \hline															
    \hline
        \\
        \end{tabular}
    \caption{Parameter estimation using Bayesian inference with datasets from different sources: original, FFNN alone, and FFNN using Monte Carlo dropout.}
    \label{tab:results}
    \end{table*}

\section{Conclusions}
\label{sec:conclusions}

    Throughout this work, we generated neural network models for cosmological datasets between redshifts $z=0$ and $z=2$ (cosmic chronometers, $f\sigma_8$ measurements, and SNeIa). We used the neural models to generate \changes{model-independent reconstructions} of $H(z)$, $f\sigma_8(z)$ and $\mu(z)$. Then, we applied Bayesian inference to \changes{data points from the reconstructions to verify that they can reproduce the expected values of the cosmological parameters} in $\Lambda$CDM and CPL models. 

    We have shown that well-calibrated artificial neural networks can produce computational models for cosmological data, even when the original datasets are small. The neural network models \changes{generate model-independent} reconstructions of the Hubble distance $H(z)$, $f\sigma_8(z)$ and distance modulus $\mu(z)$ exclusively from observational data and without assuming any cosmological model. Our results are consistent with previous works using different non-parametric inference techniques.
    
    In general, the results of the neural networks with MC-DO are better because they are considering the uncertainty of the produced models, and the dropout technique provides regularization generating a more robust model. On the other hand, the standard deviations (or variance) of the FFNN+MC-DO predictions are small, which gives us the certainty that the neural network is well-trained. The FFNN+MC-DO predictions may have more variance than the FFNN alone, and the fact that the results obtained with both models are close allows us to conclude that the FFNN predictions are acceptable.
    
    Because we are taking into account the original statistical errors as part of the training datasets, in the reconstructions of $H(z)$ and $f\sigma_8$, the errors have also been modeled by the neural networks. We are generating models for the errors; therefore, the new error bars are independent of a real data point at a given redshift, which is not the case in the Gaussian processes.
    
    {As seen in the appendices, a disadvantage of our method is that the neural networks training and their hyperparameter tuning are computationally more complex and have a higher CPU consumption than other interpolation or non-parametric inference techniques. However, our method offers some advantages that can make it a viable alternative}:
\begin{itemize}
        \item {Well-trained neural network models can be generated even with few data points.}
        \item {No fiducial cosmology is necessary to generate \changes{model-independent neural reconstructions} consistent with cosmological theory.}
        \item {No assumptions have to be made about the statistical distribution of the data.}
        \item {It allows computational models for observational data and their errors, even if they have correlations among them.}
\end{itemize}

    We have explored the generation of synthetic covariance matrices through a VAE neural network, and the results have allowed us to carry out Bayesian inference without drawbacks. The results we have obtained, as a first approach, are in agreement with other techniques. For larger datasets, we consider that using more complex architectures of autoencoders and a slightly different approach for dealing with the computing demand will be convenient.

    It is worth mentioning that the results obtained in this work are for the chosen observations and have been sufficient to show some interesting features from the data alone. In this way, we can see that using neural networks for the \changes{model-independent reconstructions} can complement the analysis of cosmological models and improve the interpretations of their behaviors. We plan to apply similar techniques to other data types, including a full set of covariance matrices, and also incorporate more sophisticated hyperparameter tuning to improve reconstructions.

\section*{Acknowledgements}

Thanks to the referees who helped improve this paper. CONACyT-Mexico Ph.D. and postdoctoral scholarships partially supported this work. RG-S acknowledges the support provided by SIP20200666-IPN and SIP20210500-IPN grants and FORDECYT-PRONACES-CONACYT CF-MG-2558591. JAV acknowledges the support provided by FOSEC SEP-CONACYT Investigaci\'on B\'asica A1-S-21925, FORDECYT-PRONACES-CONACYT 304001, and UNAM-DGAPA-PAPIIT IA104221. IGV, thanks to ICF-UNAM.

\bibliographystyle{unsrt}
\bibliography{references.bib}

\begin{thebibliography}{100}

\bibitem{copeland2006dynamics}
Edmund~J Copeland, Mohammad Sami, and Shinji Tsujikawa.
\newblock Dynamics of dark energy.
\newblock {\em Int. J. Mod. Phys. D}, 15(11):1753--1935, 2006.

\bibitem{amendola2010dark}
Luca Amendola and Shinji Tsujikawa.
\newblock {\em Dark energy: theory and observations}.
\newblock Cambridge University Press, 2010.

\bibitem{ruiz2010dark}
Pilar Ruiz-Lapuente.
\newblock {\em Dark energy: observational and theoretical approaches}.
\newblock Cambridge University Press, 2010.

\bibitem{aghanim2020planck}
Nabila Aghanim, Yashar Akrami, M~Ashdown, J~Aumont, C~Baccigalupi,
  M~Ballardini, AJ~Banday, RB~Barreiro, N~Bartolo, S~Basak, et~al.
\newblock Planck 2018 results-vi. cosmological parameters.
\newblock {\em Astronomy \& Astrophysics}, 641:A6, 2020.

\bibitem{betoule2014improved}
M~et~al Betoule, R~Kessler, J~Guy, J~Mosher, D~Hardin, R~Biswas, P~Astier,
  P~El-Hage, M~Konig, S~Kuhlmann, et~al.
\newblock Improved cosmological constraints from a joint analysis of the
  sdss-ii and snls supernova samples.
\newblock {\em Astronomy \& Astrophysics}, 568:A22, 2014.

\bibitem{stern2010cosmic}
Daniel Stern, Raul Jimenez, Licia Verde, Marc Kamionkowski, and S~Adam
  Stanford.
\newblock Cosmic chronometers: constraining the equation of state of dark
  energy. i: $h (z)$ measurements.
\newblock {\em Journal of Cosmology and Astroparticle Physics}, 2010(02):008,
  2010.

\bibitem{alam2017clustering}
Shadab Alam, Metin Ata, Stephen Bailey, Florian Beutler, Dmitry Bizyaev,
  Jonathan~A Blazek, Adam~S Bolton, Joel~R Brownstein, Angela Burden, Chia-Hsun
  Chuang, et~al.
\newblock The clustering of galaxies in the completed sdss-iii baryon
  oscillation spectroscopic survey: cosmological analysis of the dr12 galaxy
  sample.
\newblock {\em Monthly Notices of the Royal Astronomical Society},
  470(3):2617--2652, 2017.

\bibitem{sahni2002cosmological}
Varun Sahni.
\newblock The cosmological constant problem and quintessence.
\newblock {\em Class. Quantum Gravity}, 19(13):3435, 2002.

\bibitem{peebles2003cosmological}
P~James~E Peebles and Bharat Ratra.
\newblock The cosmological constant and dark energy.
\newblock {\em Rev. Modern Phys.}, 75(2):559, 2003.

\bibitem{feeney2018clarifying}
Stephen~M Feeney, Daniel~J Mortlock, and Niccolo Dalmasso.
\newblock Clarifying the hubble constant tension with a bayesian hierarchical
  model of the local distance ladder.
\newblock {\em Monthly Notices of the Royal Astronomical Society},
  476(3):3861--3882, 2018.

\bibitem{joyce2016dark}
Austin Joyce, Lucas Lombriser, and Fabian Schmidt.
\newblock Dark energy versus modified gravity.
\newblock {\em Annu. Rev. Nucl. Part. Sci.}, 66:95--122, 2016.

\bibitem{tyson2002large}
J~Anthony Tyson.
\newblock Large synoptic survey telescope: overview.
\newblock {\em Survey and Other Telescope Technologies and Discoveries},
  4836:10--20, 2002.

\bibitem{aghamousa2016desi}
Amir Aghamousa, Jessica Aguilar, Steve Ahlen, Shadab Alam, Lori~E Allen,
  Carlos~Allende Prieto, James Annis, Stephen Bailey, Christophe Balland, Otger
  Ballester, et~al.
\newblock The desi experiment part i: science, targeting, and survey design.
\newblock {\em preprint
  (\href{https://arxiv.org/abs/1611.00036}{arXiv:1611.00036})}, 2016.

\bibitem{amendola2018cosmology}
Luca Amendola, Stephen Appleby, Anastasios Avgoustidis, David Bacon, Tessa
  Baker, Marco Baldi, Nicola Bartolo, Alain Blanchard, Camille Bonvin, Stefano
  Borgani, et~al.
\newblock Cosmology and fundamental physics with the euclid satellite.
\newblock {\em Living Rev. Relativity}, 21(1):2, 2018.

\bibitem{brax2003cosmology}
Philippe Brax and Carsten van~de Bruck.
\newblock Cosmology and brane worlds: a review.
\newblock {\em Class. Quantum Gravity}, 20(9):R201, 2003.

\bibitem{clifton2012modified}
Timothy Clifton, Pedro~G Ferreira, Antonio Padilla, and Constantinos Skordis.
\newblock Modified gravity and cosmology.
\newblock {\em Physics reports}, 513(1-3):1--189, 2012.

\bibitem{vazquez2020bayesian}
J~Alberto V{\'a}zquez, David Tamayo, Anjan~A Sen, and Israel Quiros.
\newblock Bayesian model selection on scalar $\epsilon$-field dark energy.
\newblock {\em Physical Review D}, 103(4):043506, 2021.

\bibitem{quiros2018phantom}
Israel Quiros, Tame Gonzalez, Ulises Nucamendi, Ricardo Garcia-Salcedo,
  Francisco~Antonio Horta-Rangel, and Joel Saavedra.
\newblock On the phantom barrier crossing and the bounds on the speed of sound
  in non-minimal derivative coupling theories.
\newblock {\em Class. Quantum Gravity}, 35(7):075005, 2018.

\bibitem{Akarsu:2019hmw}
\"Ozg\"ur Akarsu, John~D. Barrow, Luis~A. Escamilla, and J.~Alberto Vazquez.
\newblock {Graduated dark energy: Observational hints of a spontaneous sign
  switch in the cosmological constant}.
\newblock {\em Physical Review D}, 101(6):063528, 2020.

\bibitem{chevallier2001accelerating}
Michel Chevallier and David Polarski.
\newblock Accelerating universes with scaling dark matter.
\newblock {\em Int. J. Mod. Phys. D}, 10(02):213--223, 2001.

\bibitem{sendra2012supernova}
Irene Sendra and Ruth Lazkoz.
\newblock Supernova and baryon acoustic oscillation constraints on (new)
  polynomial dark energy parametrizations: current results and forecasts.
\newblock {\em Monthly Notices of the Royal Astronomical Society},
  422(1):776--793, 2012.

\bibitem{odintsov2018cosmological}
Sergei~D Odintsov, VK~Oikonomou, AV~Timoshkin, Emmanuel~N Saridakis, and
  R~Myrzakulov.
\newblock Cosmological fluids with logarithmic equation of state.
\newblock {\em Ann. Phys.}, 398:238--253, 2018.

\bibitem{Tamayo:2019gqj}
David Tamayo and J.~Alberto Vazquez.
\newblock {Fourier-series expansion of the dark-energy equation of state}.
\newblock {\em Monthly Notices of the Royal Astronomical Society},
  487(1):729--736, 2019.

\bibitem{liu2009testing}
Jie Liu, Hong Li, Jun-Qing Xia, and Xinmin Zhang.
\newblock Testing oscillating primordial spectrum and oscillating dark energy
  with astronomical observations.
\newblock {\em Journal of Cosmology and Astroparticle Physics}, 2009(07):017,
  2009.

\bibitem{Arciniega:2021ffa}
Gustavo Arciniega, Mariana Jaber, Luisa~G Jaime, and Omar~A
  Rodr{\'\i}guez-L{\'o}pez.
\newblock One parameterisation to fit them all.
\newblock {\em preprint
  (\href{https://arxiv.org/abs/2102.08561}{arXiv:2102.08561})}, 2021.

\bibitem{Akarsu:2015yea}
\"Ozgur Akarsu, Tekin Dereli, and J.~Alberto Vazquez.
\newblock {A divergence-free parametrization for dynamical dark energy}.
\newblock {\em Journal of Cosmology and Astroparticle Physics}, 06:049, 2015.

\bibitem{wasserman2006all}
Larry Wasserman.
\newblock {\em All of nonparametric statistics}.
\newblock Springer Science \& Business Media, 2006.

\bibitem{sahni2006reconstructing}
Varun Sahni and Alexei Starobinsky.
\newblock Reconstructing dark energy.
\newblock {\em Int. J. Mod. Phys. D}, 15(12):2105--2132, 2006.

\bibitem{sharma2020reconstruction}
Ranbir Sharma, Ankan Mukherjee, and HK~Jassal.
\newblock Reconstruction of late-time cosmology using principal component
  analysis.
\newblock {\em preprint (\href{arXiv:2004.01393}{arXiv:2004.01393})}, 2020.

\bibitem{gerardi2019reconstruction}
Francesca Gerardi, Matteo Martinelli, and Alessandra Silvestri.
\newblock Reconstruction of the dark energy equation of state from latest data:
  the impact of theoretical priors.
\newblock {\em Journal of Cosmology and Astroparticle Physics}, 2019(07):042,
  2019.

\bibitem{williams2006gaussian}
Christopher~KI Williams and Carl~Edward Rasmussen.
\newblock {\em Gaussian processes for machine learning}, volume~2.
\newblock MIT press Cambridge, MA, 2006.

\bibitem{Keeley:2020aym}
Ryan~E. Keeley, Arman Shafieloo, Gong-Bo Zhao, Jose~Alberto Vazquez, and
  Hanwool Koo.
\newblock Reconstructing the universe: Testing the mutual consistency of the
  pantheon and {SDSS}/{eBOSS} {BAO} data sets with gaussian processes.
\newblock {\em The Astronomical Journal}, 161(3):151, Feb 2021.

\bibitem{l2020defying}
Benjamin L’Huillier, Arman Shafieloo, David Polarski, and Alexei~A
  Starobinsky.
\newblock Defying the laws of gravity i: model-independent reconstruction of
  the universe expansion from growth data.
\newblock {\em Monthly Notices of the Royal Astronomical Society},
  494(1):819--826, 2020.

\bibitem{mukherjee2020revisiting}
Purba Mukherjee and Narayan Banerjee.
\newblock Revisiting a non-parametric reconstruction of the deceleration
  parameter from observational data.
\newblock {\em preprint
  (\href{https://arxiv.org/abs/2007.15941}{arXiv:2007.15941})}, 2020.

\bibitem{montiel2014nonparametric}
Ariadna Montiel, Ruth Lazkoz, Irene Sendra, Celia Escamilla-Rivera, and
  Vincenzo Salzano.
\newblock Nonparametric reconstruction of the cosmic expansion with local
  regression smoothing and simulation extrapolation.
\newblock {\em Physical Review D}, 89(4):043007, 2014.

\bibitem{Vazquez:2012ce}
J.Alberto Vazquez, M.~Bridges, M.P. Hobson, and A.N. Lasenby.
\newblock {Reconstruction of the Dark Energy equation of state}.
\newblock {\em Journal of Cosmology and Astroparticle Physics}, 09:020, 2012.

\bibitem{hee2017constraining}
Sonke Hee, JA~V{\'a}zquez, WJ~Handley, MP~Hobson, and AN~Lasenby.
\newblock Constraining the dark energy equation of state using bayes theorem
  and the kullback--leibler divergence.
\newblock {\em Monthly Notices of the Royal Astronomical Society},
  466(1):369--377, 2017.

\bibitem{holsclaw2010nonparametric}
Tracy Holsclaw, Ujjaini Alam, Bruno Sanso, Herbert Lee, Katrin Heitmann, Salman
  Habib, and David Higdon.
\newblock Nonparametric dark energy reconstruction from supernova data.
\newblock {\em Physical Review Letters}, 105(24):241302, 2010.

\bibitem{zhao2017dynamical}
Gong-Bo Zhao, Marco Raveri, Levon Pogosian, Yuting Wang, Robert~G Crittenden,
  Will~J Handley, Will~J Percival, Florian Beutler, Jonathan Brinkmann,
  Chia-Hsun Chuang, et~al.
\newblock Dynamical dark energy in light of the latest observations.
\newblock {\em Nature Astronomy}, 1(9):627--632, 2017.

\bibitem{wei2017improved}
Jun-Jie Wei and Xue-Feng Wu.
\newblock An improved method to measure the cosmic curvature.
\newblock {\em The Astrophysical Journal}, 838(2):160, 2017.

\bibitem{lin2019non}
Hai-Nan Lin, Xin Li, and Li~Tang.
\newblock Non-parametric reconstruction of dark energy and cosmic expansion
  from the pantheon compilation of type ia supernovae.
\newblock {\em Chinese Phys. C}, 43(7):075101, 2019.

\bibitem{Vazquez:2012ux}
J.Alberto Vazquez, M.~Bridges, M.P. Hobson, and A.N. Lasenby.
\newblock {Model selection applied to reconstruction of the Primordial Power
  Spectrum}.
\newblock {\em Journal of Cosmology and Astroparticle Physics}, 06:006, 2012.

\bibitem{Handley:2019fll}
Will~J. Handley, Anthony~N. Lasenby, Hiranya~V. Peiris, and Michael~P. Hobson.
\newblock {Bayesian inflationary reconstructions from Planck 2018 data}.
\newblock {\em Physical Review D}, 100(10):103511, 2019.

\bibitem{lin2017does}
Henry~W Lin, Max Tegmark, and David Rolnick.
\newblock Why does deep and cheap learning work so well?
\newblock {\em J. Statistical Physics}, 168(6):1223--1247, 2017.

\bibitem{peel2019distinguishing}
Austin Peel, Florian Lalande, Jean-Luc Starck, Valeria Pettorino, Julian
  Merten, Carlo Giocoli, Massimo Meneghetti, and Marco Baldi.
\newblock Distinguishing standard and modified gravity cosmologies with machine
  learning.
\newblock {\em Physical Review D}, 100(2):023508, 2019.

\bibitem{arjona2020can}
Rub{\'e}n Arjona and Savvas Nesseris.
\newblock What can machine learning tell us about the background expansion of
  the universe?
\newblock {\em Physical Review D}, 101(12):123525, 2020.

\bibitem{wang2020machine}
Guo-Jian Wang, Xiao-Jiao Ma, and Jun-Qing Xia.
\newblock Machine learning the cosmic curvature in a model-independent way.
\newblock {\em Monthly Notices of the Royal Astronomical Society},
  501(4):5714--5722, 2021.

\bibitem{gomez2021neural}
Isidro G{\'o}mez-Vargas, Ricardo~Medel Esquivel, Ricardo Garc{\'\i}a-Salcedo,
  and J~Alberto V{\'a}zquez.
\newblock Neural network within a bayesian inference framework.
\newblock {\em J. Phys. Conf. Ser.}, 1723(1):012022, 2021.

\bibitem{Chacon:2021sil}
Jazhiel Chac\'on, J.~Alberto V\'azquez, and Erick Almaraz.
\newblock {Classification algorithms applied to structure formation
  simulations}.
\newblock {\em \href{https://arxiv.org/abs/2106.06587}{arXiv:2106.06587}}, 6
  2021.

\bibitem{dieleman2015rotation}
Sander Dieleman, Kyle~W Willett, and Joni Dambre.
\newblock Rotation-invariant convolutional neural networks for galaxy
  morphology prediction.
\newblock {\em Monthly Notices of the Royal Astronomical Society},
  450(2):1441--1459, 2015.

\bibitem{ntampaka2019deep}
Michelle Ntampaka, J~ZuHone, D~Eisenstein, D~Nagai, A~Vikhlinin, L~Hernquist,
  F~Marinacci, D~Nelson, R~Pakmor, A~Pillepich, et~al.
\newblock A deep learning approach to galaxy cluster x-ray masses.
\newblock {\em The Astrophysical Journal}, 876(1):82, 2019.

\bibitem{rodriguez2018fast}
Andres~C Rodr{\'\i}guez, Tomasz Kacprzak, Aurelien Lucchi, Adam Amara, Raphael
  Sgier, Janis Fluri, Thomas Hofmann, and Alexandre R{\'e}fr{\'e}gier.
\newblock Fast cosmic web simulations with generative adversarial networks.
\newblock {\em Comp. Astrophys. and Cosmology}, 5(1):4, 2018.

\bibitem{he2019learning}
Siyu He, Yin Li, Yu~Feng, Shirley Ho, Siamak Ravanbakhsh, Wei Chen, and
  Barnab{\'a}s P{\'o}czos.
\newblock Learning to predict the cosmological structure formation.
\newblock {\em Proc. Natl. Acad. Sci.}, 116(28):13825--13832, 2019.

\bibitem{auld2007fast}
T~Auld, Michael Bridges, MP~Hobson, and SF~Gull.
\newblock Fast cosmological parameter estimation using neural networks.
\newblock {\em Monthly Notices of the Royal Astronomical Society: Letters},
  376(1):L11--L15, 2007.

\bibitem{alsing2019fast}
Justin Alsing, Tom Charnock, Stephen Feeney, and Benjamin Wandelt.
\newblock Fast likelihood-free cosmology with neural density estimators and
  active learning.
\newblock {\em Monthly Notices of the Royal Astronomical Society},
  488(3):4440--4458, 2019.

\bibitem{li2019model}
Shi-Yu Li, Yun-Long Li, and Tong-Jie Zhang.
\newblock Model comparison of dark energy models using deep network.
\newblock {\em Res. Astron. Astrophys.}, 19(9):137, 2019.

\bibitem{hortua2020constraining}
H{\'e}ctor~J Hort{\'u}a, Luigi Malag{\`o}, and Riccardo Volpi.
\newblock Constraining the reionization history using bayesian normalizing
  flows.
\newblock {\em Machine Learning: Science and Technology}, 1(3):035014, 2020.

\bibitem{hortua2020parameter}
H{\'e}ctor~J Hort{\'u}a, Riccardo Volpi, Dimitri Marinelli, and Luigi
  Malag{\`o}.
\newblock Parameter estimation for the cosmic microwave background with
  bayesian neural networks.
\newblock {\em Physical Review D}, 102(10):103509, 2020.

\bibitem{escamilla2020deep}
Celia Escamilla-Rivera, Maryi A~Carvajal Quintero, and Salvatore Capozziello.
\newblock A deep learning approach to cosmological dark energy models.
\newblock {\em Journal of Cosmology and Astroparticle Physics}, 2020(03):008,
  2020.

\bibitem{wang2020reconstructing}
Guo-Jian Wang, Xiao-Jiao Ma, Si-Yao Li, and Jun-Qing Xia.
\newblock Reconstructing functions and estimating parameters with artificial
  neural networks: A test with a hubble parameter and sne ia.
\newblock {\em Astrophysical Journal Supplement Series}, 246(1):13, 2020.

\bibitem{dialektopoulos2021neural}
Konstantinos Dialektopoulos, Jackson~Levi Said, Jurgen Mifsud, Joseph Sultana,
  and Kristian~Zarb Adami.
\newblock Neural network reconstruction of late-time cosmology and null tests.
\newblock {\em arXiv preprint arXiv:2111.11462}, 2021.

\bibitem{linder2003exploring}
Eric~V Linder.
\newblock Exploring the expansion history of the universe.
\newblock {\em Physical Review Letters}, 90(9):091301, 2003.

\bibitem{jimenez2003constraints}
Raul Jimenez, Licia Verde, Tommaso Treu, and Daniel Stern.
\newblock Constraints on the equation of state of dark energy and the hubble
  constant from stellar ages and the cosmic microwave background.
\newblock {\em The Astrophysical Journal}, 593(2):622, 2003.

\bibitem{simon2005constraints}
Joan Simon, Licia Verde, and Raul Jimenez.
\newblock Constraints on the redshift dependence of the dark energy potential.
\newblock {\em Physical Review D}, 71(12):123001, 2005.

\bibitem{moresco2012new}
Michele Moresco, Licia Verde, Lucia Pozzetti, Raul Jimenez, and Andrea Cimatti.
\newblock New constraints on cosmological parameters and neutrino properties
  using the expansion rate of the universe to $z \sim 1.75$.
\newblock {\em Journal of Cosmology and Astroparticle Physics}, 2012(07):053,
  2012.

\bibitem{zhang2014four}
Cong Zhang, Han Zhang, Shuo Yuan, Siqi Liu, Tong-Jie Zhang, and Yan-Chun Sun.
\newblock Four new observational $h (z)$ data from luminous red galaxies in the
  sloan digital sky survey data release seven.
\newblock {\em Res. Astron. Astrophys.}, 14(10):1221, 2014.

\bibitem{moresco2015raising}
Michele Moresco.
\newblock Raising the bar: new constraints on the hubble parameter with cosmic
  chronometers at $z \sim 2$.
\newblock {\em Monthly Notices of the Royal Astronomical Society: Letters},
  450(1):L16--L20, 2015.

\bibitem{moresco20166}
Michele Moresco, Lucia Pozzetti, Andrea Cimatti, Raul Jimenez, Claudia
  Maraston, Licia Verde, Daniel Thomas, Annalisa Citro, Rita Tojeiro, and David
  Wilkinson.
\newblock A 6\% measurement of the hubble parameter at $z \sim 0.45$: direct
  evidence of the epoch of cosmic re-acceleration.
\newblock {\em Journal of Cosmology and Astroparticle Physics}, 2016(05):014,
  2016.

\bibitem{ratsimbazafy2017age}
AL~Ratsimbazafy, SI~Loubser, SM~Crawford, CM~Cress, BA~Bassett, RC~Nichol, and
  P~V{\"a}is{\"a}nen.
\newblock Age-dating luminous red galaxies observed with the southern african
  large telescope.
\newblock {\em Monthly Notices of the Royal Astronomical Society},
  467(3):3239--3254, 2017.

\bibitem{said2020joint}
Khaled Said, Matthew Colless, Christina Magoulas, John~R Lucey, and Michael~J
  Hudson.
\newblock Joint analysis of 6dfgs and sdss peculiar velocities for the growth
  rate of cosmic structure and tests of gravity.
\newblock {\em Monthly Notices of the Royal Astronomical Society},
  497(1):1275--1293, 2020.

\bibitem{kaiser1987clustering}
Nick Kaiser.
\newblock Clustering in real space and in redshift space.
\newblock {\em Monthly Notices of the Royal Astronomical Society},
  227(1):1--21, 1987.

\bibitem{amendola2008measuring}
Luca Amendola, Martin Kunz, and Domenico Sapone.
\newblock Measuring the dark side (with weak lensing).
\newblock {\em Journal of Cosmology and Astroparticle Physics}, 2008(04):013,
  2008.

\bibitem{sagredo2018internal}
Bryan Sagredo, Savvas Nesseris, and Domenico Sapone.
\newblock Internal robustness of growth rate data.
\newblock {\em Physical Review D}, 98(8):083543, 2018.

\bibitem{trotta2008bayes}
Roberto Trotta.
\newblock Bayes in the sky: Bayesian inference and model selection in
  cosmology.
\newblock {\em Contemporary Phys.}, 49(2):71--104, 2008.

\bibitem{feroz2009}
F~Feroz, MP~Hobson, and M~Bridges.
\newblock Multinest: an efficient and robust bayesian inference tool for
  cosmology and particle physics.
\newblock {\em Monthly Notices of the Royal Astronomical Society},
  398(4):1601--1614, 2009.

\bibitem{leclercq2018bayesian}
Florent Leclercq.
\newblock Bayesian optimization for likelihood-free cosmological inference.
\newblock {\em Physical Review D}, 98(6):063511, 2018.

\bibitem{taylor2010analytic}
AN~Taylor and TD~Kitching.
\newblock Analytic methods for cosmological likelihoods.
\newblock {\em Monthly Notices of the Royal Astronomical Society},
  408(2):865--875, 2010.

\bibitem{nesseris2012new}
Savvas Nesseris and Juan Garcia-Bellido.
\newblock A new perspective on dark energy modeling via genetic algorithms.
\newblock {\em Journal of Cosmology and Astroparticle Physics}, 2012(11):033,
  2012.

\bibitem{hannestad1999stochastic}
Steen Hannestad.
\newblock Stochastic optimization methods for extracting cosmological
  parameters from cosmic microwave background radiation power spectra.
\newblock {\em Physical Review D}, 61(2):023002, 1999.

\bibitem{prasad2012cosmological}
Jayanti Prasad and Tarun Souradeep.
\newblock Cosmological parameter estimation using particle swarm optimization.
\newblock {\em Physical Review D}, 85(12):123008, 2012.

\bibitem{padilla2019}
Luis~E Padilla, Luis~O Tellez, Luis~A Escamilla, and Jose~Alberto Vazquez.
\newblock Cosmological parameter inference with bayesian statistics.
\newblock {\em Universe}, 7(7):213, 2021.

\bibitem{medel2021}
Ricardo Medel~Esquivel, Isidro G{\'o}mez-Vargas, J~Alberto V{\'a}zquez, and
  Ricardo García~Salcedo.
\newblock An introduction to markov chain monte carlo.
\newblock {\em
  \href{http://www.seio.es/BBEIO/BEIOVol37Num1/files/assets/basic-html/page-51.html}{Bolet{\'\i}n
  de Estad{\'\i}stica e Investigaci{\'o}n Operativa}}, 1(37):47--74, 2021.

\bibitem{juanUniverse}
Juan de~Dios Rojas~Olvera, Isidro G{\'o}mez-Vargas, and Jose~Alberto
  V{\'a}zquez.
\newblock Observational cosmology with artificial neural networks.
\newblock {\em Universe}, 8(2):120, 2022.

\bibitem{hornik1990universal}
Kurt Hornik, Maxwell Stinchcombe, and Halbert White.
\newblock Universal approximation of an unknown mapping and its derivatives
  using multilayer feedforward networks.
\newblock {\em Neural networks}, 3(5):551--560, 1990.

\bibitem{goodfellow2016deep}
Ian Goodfellow, Yoshua Bengio, Aaron Courville, and Yoshua Bengio.
\newblock {\em Deep learning}, volume~1.
\newblock MIT press Cambridge, 2016.

\bibitem{baldi1989neural}
Pierre Baldi and Kurt Hornik.
\newblock Neural networks and principal component analysis: Learning from
  examples without local minima.
\newblock {\em Neural networks}, 2(1):53--58, 1989.

\bibitem{ingrassia2005neural}
Salvatore Ingrassia and Isabella Morlini.
\newblock Neural network modeling for small datasets.
\newblock {\em Technometrics}, 47(3):297--311, 2005.

\bibitem{ng2015deep}
Hong-Wei Ng, Viet~Dung Nguyen, Vassilios Vonikakis, and Stefan Winkler.
\newblock Deep learning for emotion recognition on small datasets using
  transfer learning.
\newblock In {\em Proceedings of the 2015 ACM on international conference on
  multimodal interaction}, pages 443--449, 2015.

\bibitem{pasini2015artificial}
Antonello Pasini.
\newblock Artificial neural networks for small dataset analysis.
\newblock {\em Journal of thoracic disease}, 7(5):953, 2015.

\bibitem{wang2020generalizing}
Yaqing Wang, Quanming Yao, James~T Kwok, and Lionel~M Ni.
\newblock Generalizing from a few examples: A survey on few-shot learning.
\newblock {\em ACM Computing Surveys (CSUR)}, 53(3):1--34, 2020.

\bibitem{gal2015dropout}
Yarin Gal and Zoubin Ghahramani.
\newblock Dropout as a bayesian approximation: Insights and applications.
\newblock {\em Deep Learning Workshop, ICML}, 1:2, 2015.

\bibitem{riess20162}
Adam~G Riess, Lucas~M Macri, Samantha~L Hoffmann, Dan Scolnic, Stefano
  Casertano, Alexei~V Filippenko, Brad~E Tucker, Mark~J Reid, David~O Jones,
  Jeffrey~M Silverman, et~al.
\newblock A 2.4\% determination of the local value of the hubble constant.
\newblock {\em The Astrophysical Journal}, 826(1):56, 2016.

\bibitem{singirikonda2020model}
Haveesh Singirikonda and Shantanu Desai.
\newblock Model comparison of $\lambda$cdm vs $r_h= ct $ using cosmic
  chronometers.
\newblock {\em The European Physical Journal C}, 80(8):1--9, 2020.

\bibitem{mukherjee2021assessment}
Purba Mukherjee and Ankan Mukherjee.
\newblock Assessment of the cosmic distance duality relation using gaussian
  process.
\newblock {\em Monthly Notices of the Royal Astronomical Society},
  504(3):3938--3946, 2021.

\bibitem{bonilla2021measurements}
Alexander Bonilla, Suresh Kumar, and Rafael~C Nunes.
\newblock Measurements of $h\_0$ and reconstruction of the dark energy
  properties from a model-independent joint analysis.
\newblock {\em The European Physical Journal C}, 81(2):1--13, 2021.

\bibitem{ma2011power}
Cong Ma and Tong-Jie Zhang.
\newblock Power of observational hubble parameter data: a figure of merit
  exploration.
\newblock {\em The Astrophysical Journal}, 730(2):74, 2011.

\bibitem{yin2019non}
Zhao-Yu Yin and Hao Wei.
\newblock Non-parametric reconstruction of growth index via gaussian processes.
\newblock {\em SCIENCE CHINA Physics, Mechanics \& Astronomy}, 62(9):1--10,
  2019.

\bibitem{rumelhart1986learning}
David~E Rumelhart, Geoffrey~E Hinton, and Ronald~J Williams.
\newblock Learning representations by back-propagating errors.
\newblock {\em Nature}, 323(6088):533--536, 1986.

\bibitem{lecun2012efficient}
Yann~A LeCun, L{\'e}on Bottou, Genevieve~B Orr, and Klaus-Robert M{\"u}ller.
\newblock Efficient backprop.
\newblock {\em Neural networks: Tricks of the trade}, pages 9--48, 2012.

\bibitem{srivastava2014dropout}
Nitish Srivastava, Geoffrey Hinton, Alex Krizhevsky, Ilya Sutskever, and Ruslan
  Salakhutdinov.
\newblock Dropout: a simple way to prevent neural networks from overfitting.
\newblock {\em J. Machine Learning Res.}, 15(1):1929--1958, 2014.

\bibitem{kingma2013auto}
Diederik~P Kingma and Max Welling.
\newblock Auto-encoding variational bayes.
\newblock {\em preprint
  (\href{https://arxiv.org/abs/1312.6114}{arXiv:1312.6114)}}, 2013.

\bibitem{rezende2014stochastic}
Danilo~Jimenez Rezende, Shakir Mohamed, and Daan Wierstra.
\newblock Stochastic backpropagation and approximate inference in deep
  generative models.
\newblock {\em \href{https://proceedings.mlr.press/v32/rezende14.html}{Int.
  Conf. Machine Learning}}, pages 1278--1286, 2014.

\bibitem{kullback1951information}
Solomon Kullback and Richard~A Leibler.
\newblock On information and sufficiency.
\newblock {\em The Annals of Mathematical Statistics}, 22(1):79--86, 1951.

\bibitem{ramos2018binary}
Daniel Ramos, Javier Franco-Pedroso, Alicia Lozano-Diez, and Joaquin
  Gonzalez-Rodriguez.
\newblock Deconstructing cross-entropy for probabilistic binary classifiers.
\newblock {\em Entropy}, 20(3):208, 2018.

\bibitem{doersch2016tutorial}
Carl Doersch.
\newblock Tutorial on variational autoencoders.
\newblock {\em preprint
  (\href{https://arxiv.org/abs/1606.05908}{arXiv:1606.05908})}, 2016.

\bibitem{kingma2019introduction}
Diederik~P Kingma and Max Welling.
\newblock An introduction to variational autoencoders.
\newblock {\em preprint
  (\href{https://arxiv.org/abs/1906.02691}{arXiv:1906.02691})}, 2019.

\bibitem{geman1992neural}
Stuart Geman, Elie Bienenstock, and Ren{\'e} Doursat.
\newblock Neural networks and the bias/variance dilemma.
\newblock {\em Neural Comp.}, 4(1):1--58, 1992.

\bibitem{larochelle2007empirical}
Hugo Larochelle, Dumitru Erhan, Aaron Courville, James Bergstra, and Yoshua
  Bengio.
\newblock An empirical evaluation of deep architectures on problems with many
  factors of variation.
\newblock {\em Proc. 24th Int. Conf. Machine Learning}, pages 473--480, 2007.

\bibitem{hutter2009paramils}
Frank Hutter, Holger~H Hoos, Kevin Leyton-Brown, and Thomas St{\"u}tzle.
\newblock Paramils: an automatic algorithm configuration framework.
\newblock {\em J. Artif. Intell. Res.}, 36:267--306, 2009.

\bibitem{bardenet2013collaborative}
Rémi Bardenet, Mátyás Brendel, Balázs Kégl, and Michèle Sebag.
\newblock Collaborative hyperparameter tuning.
\newblock {\em Proc. 30th Int. Conf. Machine Learning}, 28(2):199--207, 2013.

\bibitem{zhang2019deep}
Xiang Zhang, Xiaocong Chen, Lina Yao, Chang Ge, and Manqing Dong.
\newblock Deep neural network hyperparameter optimization with orthogonal array
  tuning.
\newblock {\em Int. Conf. Neural Inf. Processing}, pages 287--295, 2019.

\bibitem{aubourg2015}
{\'E}ric Aubourg, Stephen Bailey, Julian~E Bautista, Florian Beutler, Vaishali
  Bhardwaj, Dmitry Bizyaev, Michael Blanton, Michael Blomqvist, Adam~S Bolton,
  Jo~Bovy, et~al.
\newblock Cosmological implications of baryon acoustic oscillation
  measurements.
\newblock {\em Physical Review D}, 92(12):123516, 2015.

\bibitem{speagle2020dynesty}
Joshua~S Speagle.
\newblock dynesty: a dynamic nested sampling package for estimating bayesian
  posteriors and evidences.
\newblock {\em Monthly Notices of the Royal Astronomical Society},
  493(3):3132--3158, 2020.

\bibitem{leung2019deep}
Henry~W Leung and Jo~Bovy.
\newblock Deep learning of multi-element abundances from high-resolution
  spectroscopic data.
\newblock {\em Monthly Notices of the Royal Astronomical Society},
  483(3):3255--3277, 2019.

\end{thebibliography}

\appendix 
\section{{Neural Networks basics}}
 \label{sec:appendix_ann_basics}
    {Here, we present the learning mechanism of an ANN with some of the settings we have used throughout this work}:
 \begin{itemize}
        \item {Before the ANN training, we split the original datasets into training and validation sets with $80\%$ and $20\%$, respectively.} The first set is used to train the ANN, while the validation set contains unseen values. Therefore, it helps test the performance of the ANN and evaluate its ability to produce an excellent model for the input dataset.    
        \item The first layer of neurons reads the dataset’s features (or columns). Each connection between neurons is assigned a random number called weight (we use random numbers with a normal distribution centered on $0$ with a standard deviation of $0.01$). The input data make up a matrix $X_1$ and provide the values for the first layer of nodes. The $X_i$ refers to the values of nodes in the $i$-th  layer. The weights make up another matrix $W_i$, which are the values for the connections between the $i$-th and the $(i+1)$-th layers. In addition, each connection has a bias term $b_i$. The product $Z$ of these two matrices plus the bias is as follows:

        \begin{equation}
            Z_{i+1} = W_i^T X_i+b_i,
        \end{equation}
        where $W_i \in  \mathbb{R}^{m \times n}$, with $m$ and $n$ the number of nodes in the $i$-th and $(i+1)$-th layers respectively. $X_i$ corresponds to the $i$-th layer, therefore, has $m$ dimensions. It is worth applying the transpose of $W_i$ to allow the matrix product.
           
        \item An activation (or transfer) function $\phi$ modulates $Z_i$ and assigns values to the next layer of neurons. This process, known as forward propagation, is repeated until the last layer is reached. The values of neurons in subsequent layers are given by:

        \begin{equation}
            X_{i+1} = \phi(Z_{i+1}).
        \end{equation}
        \item The value of the neurons in the last layer must be evaluated by an error function (or loss function) which measures the difference between the value given by the ANN and the expected one. {In order to find the better values of weights,} the loss function is minimized by an optimization method such as \textit{gradient descent} combined with the \textit{backpropagation} algorithm to compute gradients \cite{rumelhart1986learning, lecun2012efficient}. 
     
        \item During backpropagation, the weights are updated, then forward propagation is performed again. This is repeated until the loss function reaches the desired precision, and then the neural network is trained and ready to make predictions. The number of samples propagated through the network before updating the weights is known as \textit{batch size}, and each iteration of the entire dataset constitutes an \textit{epoch}.

    \end{itemize}
    
    Another essential concept is the dropout (DO), a regularization technique \cite{srivastava2014dropout} that allows smaller values to be achieved in the loss function and prevents overfitting. It consists in randomly turning off neurons during training, so the neurons that operate at each epoch are different. The associated hyperparameter is a scalar value that indicates the probability of turning off a neuron in each epoch. Due to its random nature, the dropout can be used as a Monte Carlo simulation. When an ANN is trained, the dropout can be interpreted as a Bayesian approximation of a gaussian probabilistic model, during training the active neurons are different, hence different weight configurations are saved, and in each prediction, the neural network uses a different one, such as in each case we use a different neural network, with other neurons turned off. Therefore, it is possible to make several predictions and thus obtain the average and standard deviations. Using this formalism, dubbed Monte Carlo dropout (MC-DO) \cite{gal2015dropout}, we can get a statistical uncertainty of a trained ANN model. We apply the dropout method to the FFNNs implemented in this work and compare the results with those solely with FFNNs.
    
    The intrinsic parameters of an ANN are known as hyperparameters: the number of layers, number of nodes, batch size, dropout value, optimizer algorithm, or number of epochs, among others. It is worth carefully selecting a good combination of them to guarantee that the ANN model has the capability of generalization. An incorrect choice of them can produce undesirable models, either underfitted or overfitted concerning the data. 
    
\section{Autoencoders}
\label{sec:appendix_autoencoders}

    Autoencoders can be thought as two symmetrical coupled ANNs, where the first (encoder) makes a dimensional reduction for the input and obtains a coded representation (vector embedding or latent space) of the original data. The second part (decoder) takes the coded representation of the data and recovers an instance with the same data type and dimension as the original input. The encoder is a function $f$ that maps the input $x$ with dimension $l$ to an encoded vector $h$ with dimension $m$, with $m < l$: 
    \begin{equation}
        f : x \in \mathbb{R}^l \rightarrow h \in \mathbb{R}^m,
        \label{eq:encoder}
    \end{equation}
    where $h_i := f_i(x) = \phi_e(W^T_i X_i +b^e_i), \; i=1,2,...,m$ with $\phi$ being the activation function and the $e$ index refers to the encoder. The decoder is the following $g$ function that maps the encoded representation with dimension $m$ into an output $\hat{x}$ with the same dimension $l$ as the original input $x$: 
         \begin{equation}
             g: h \in \mathbb{R}^m \rightarrow \hat{x} \in \mathbb{R}^l,
             \label{eq:decoder}
         \end{equation}
        \noindent
    with $\hat{x}_j = g_j(h)=\phi_d(W'_jh+b^d_j), \; i=1,2,...,l$, where the $d$ index refers to the decoder. The goal of the autoencoder training is to find the parameters of the functions shown in equations \ref{eq:encoder} and \ref{eq:decoder}: $[W_1, ..., W_m], b^e$ for the encoder and $[W'_1, ..., W'_m], b^d$ for the decoder. If the activation function is the identity function, \textit{i.e.}, $\phi(x) = x$, then this type of neural network is analogous to the Principal Component Analysis (PCA) technique. In this work, we use a particular type called variational autoencoder (VAE), which belongs to the so-called generative neural networks  \cite{kingma2013auto, rezende2014stochastic}.
    
    VAE neural networks use variational inference to sample the compressed representation (or latent space) and, therefore, allow us to know the probability function associated precisely, with the compressed representation. Unlike classical autoencoders, such as those described earlier in this work, two layers of the same dimension as the latent space are designed before the compressed representation, whose function is to generate values to sample the mean $\mu$ and variance $\sigma$, which are the parameters of the statistical distribution that produces an input data (matrix or image) of the VAE to generate a point $z$ of the latent space. 

    To construct a latent space distribution similar to the proposed Gaussian distribution, the Kullback-Leiber divergence (KL) is used \cite{kullback1951information}. Thus, the selection of the relevant loss function to train the VAE is as follows: 
    \begin{equation}\label{eq:lossvae}
        loss_{\rm VAE} = {\rm MSE} + \mathbf{K} \mathbf{L}(q(z|x)||p(z)),
    \end{equation}
  where $q(z|x)$ is the probability density function to generate a $z$ point of the latent space given an input $x$. On the other hand, we can assume that $p(z) =N(0, I)$ with $p$ a probability density function of the $z$ points in latent space and $N$ a normal distribution centered at $0$ with covariance matrix equal to the identity matrix. Because VAEs are widely used in image processing, it is more common to choose \textit{binary cross entropy} \cite{ramos2018binary} instead of MSE. However, our interest is in the numerical information of the covariance matrices and not just in a classification problem in image generation. 
  
  For an extended review about Variational Autoencoders, we recommend \cite{doersch2016tutorial} and \cite{kingma2019introduction}.
    
\section{{Feedforward neural networks training}}
\label{sec:appendix_ffnn_train}
    \begin{figure*}
    \centering
    \captionsetup{justification=raggedright,singlelinecheck=false,font=footnotesize}
     \makebox[12cm][c]{
            \includegraphics[trim= 10mm 20mm 20mm 15mm, clip, width=4.5cm, height=3.cm]{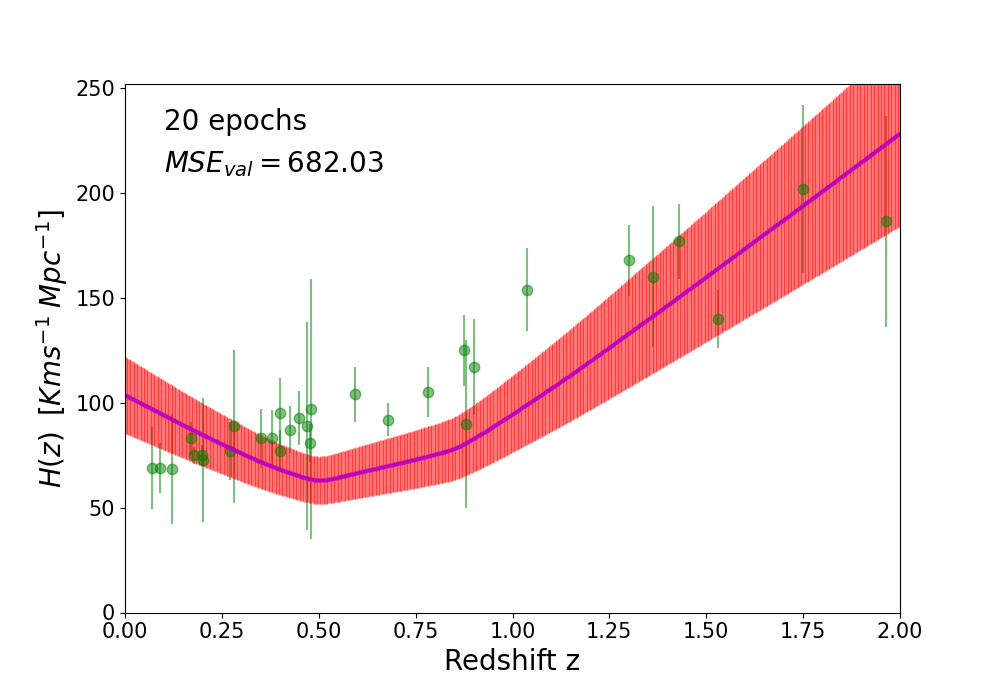}  
            \includegraphics[trim= 31mm 20mm 20mm 15mm, clip, width=4.3cm, height=3.cm]{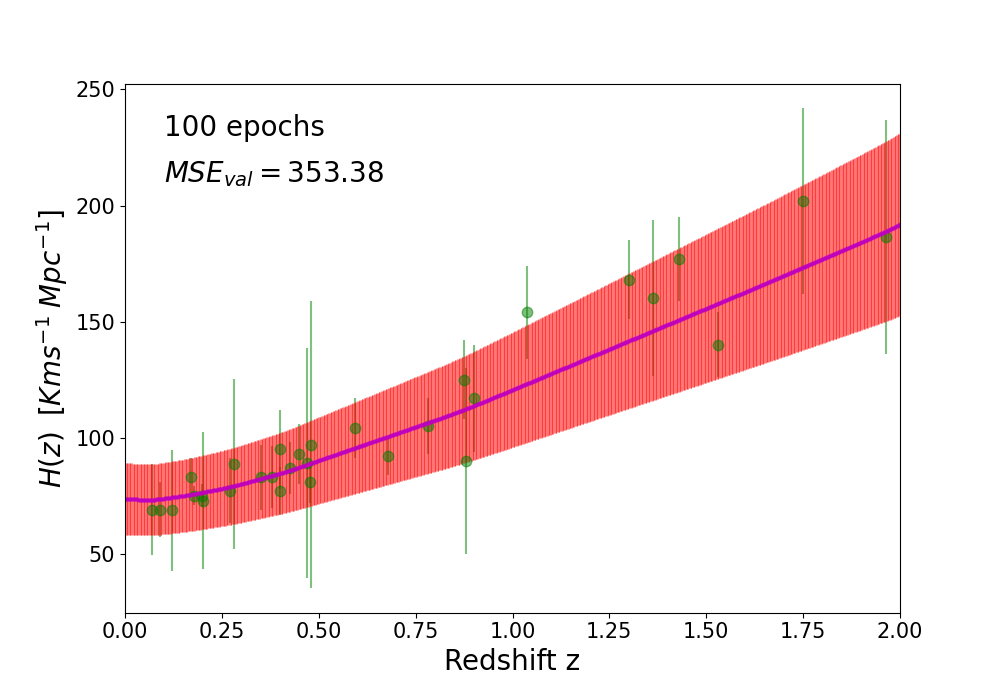}  
            \includegraphics[trim= 31mm 20mm 20mm 15mm, clip, width=4.3cm, height=3.cm]{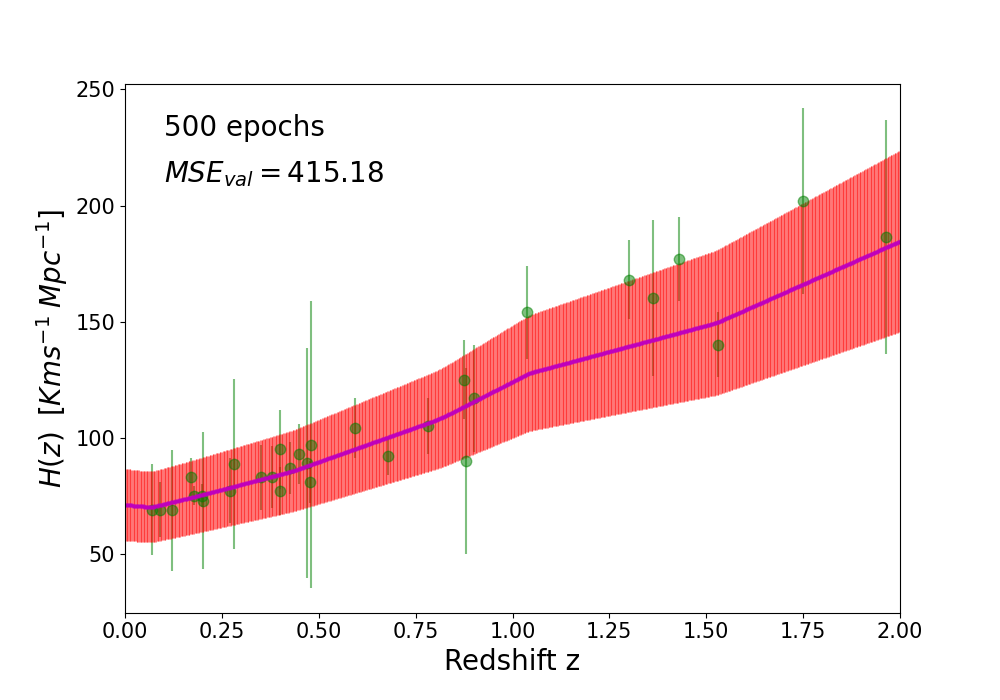}  
            \includegraphics[trim= 31mm 20mm 20mm 15mm, clip, width=4.3cm, height=3.cm]{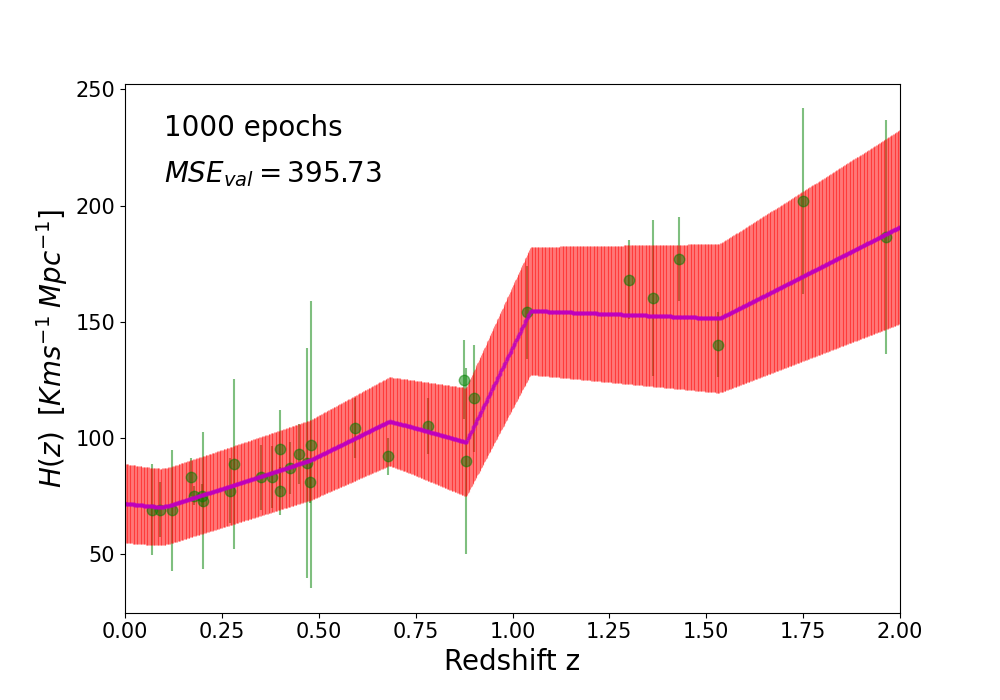} 
            }
    \makebox[12cm][c]{
            \includegraphics[trim= 10mm 0mm 20mm 15mm, clip, width=4.5cm, height=3.cm]{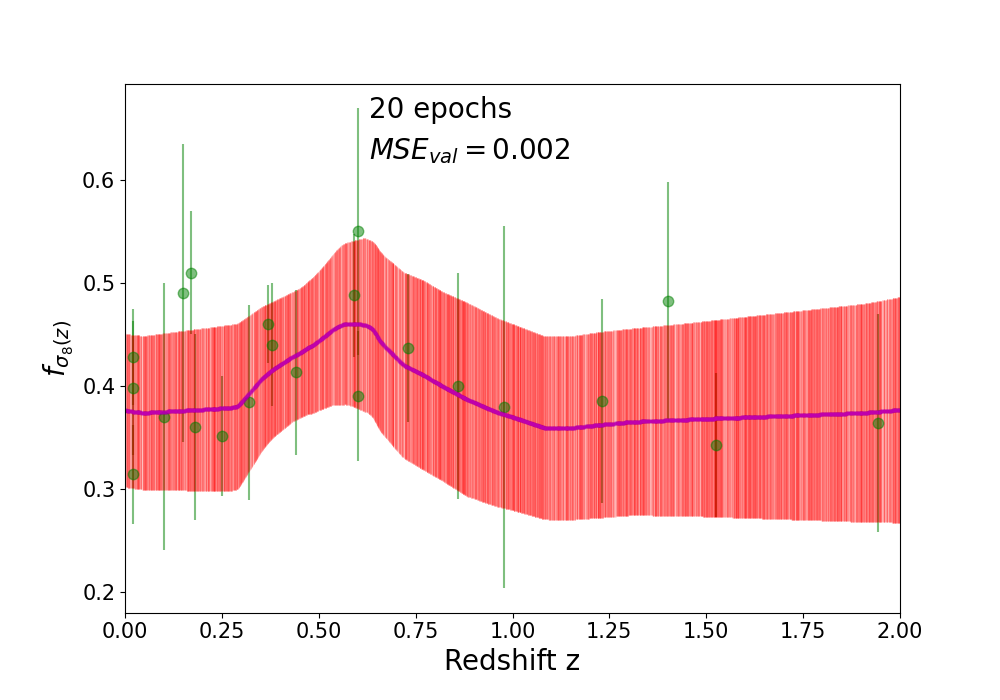}  
            \includegraphics[trim= 31mm 0mm 20mm 15mm, clip, width=4.3cm, height=3.cm]{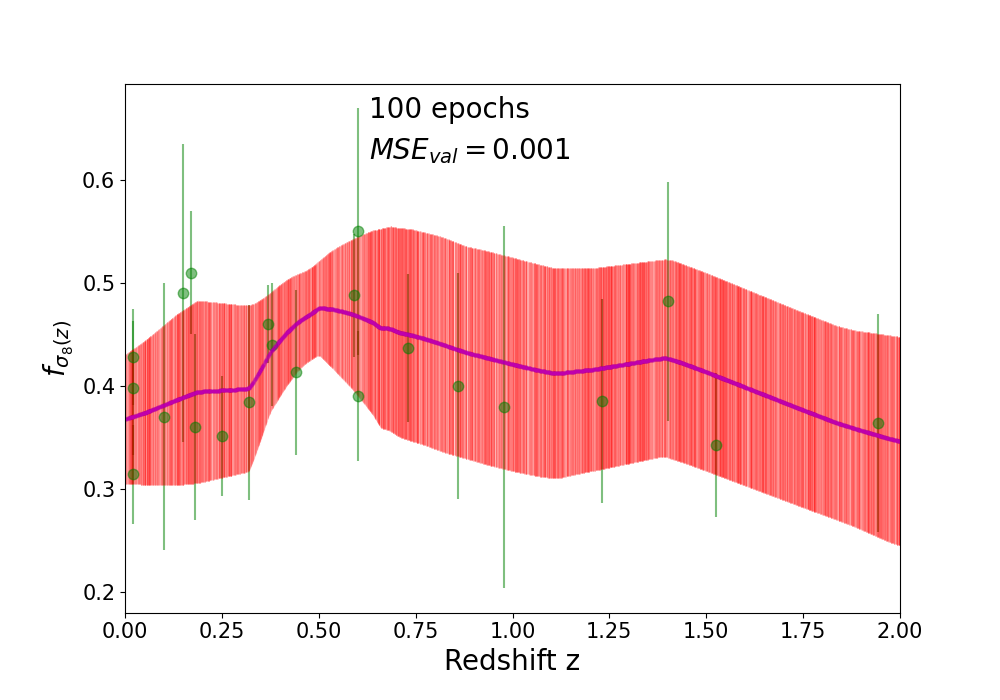}  
            \includegraphics[trim= 31mm 0mm 20mm 15mm, clip, width=4.3cm, height=3.cm]{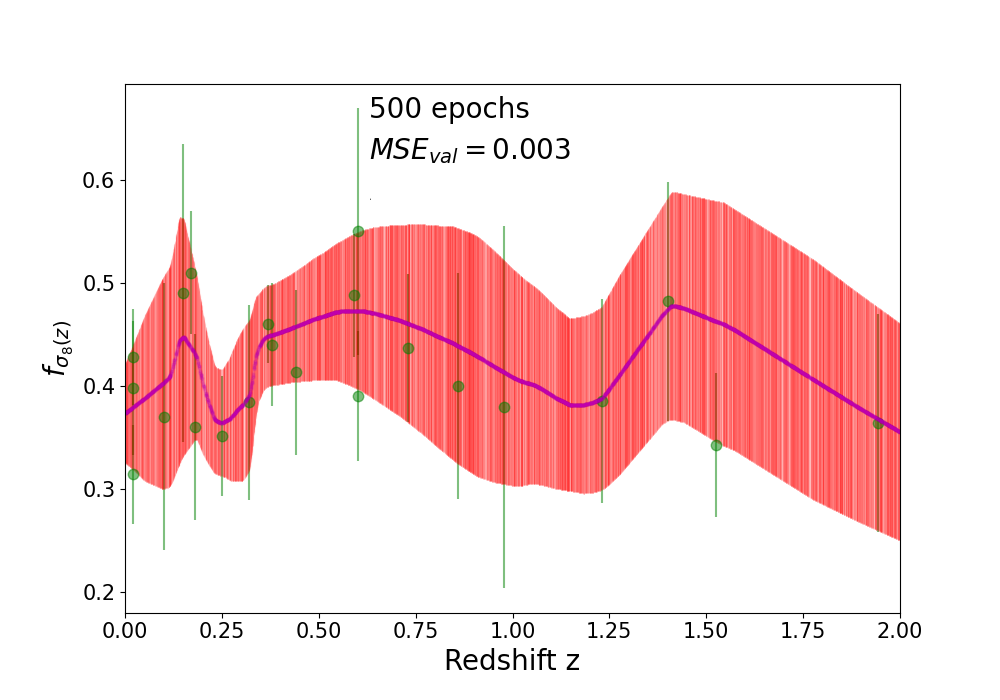}  
            \includegraphics[trim= 31mm 0mm 20mm 15mm, clip, width=4.3cm, height=3.cm]{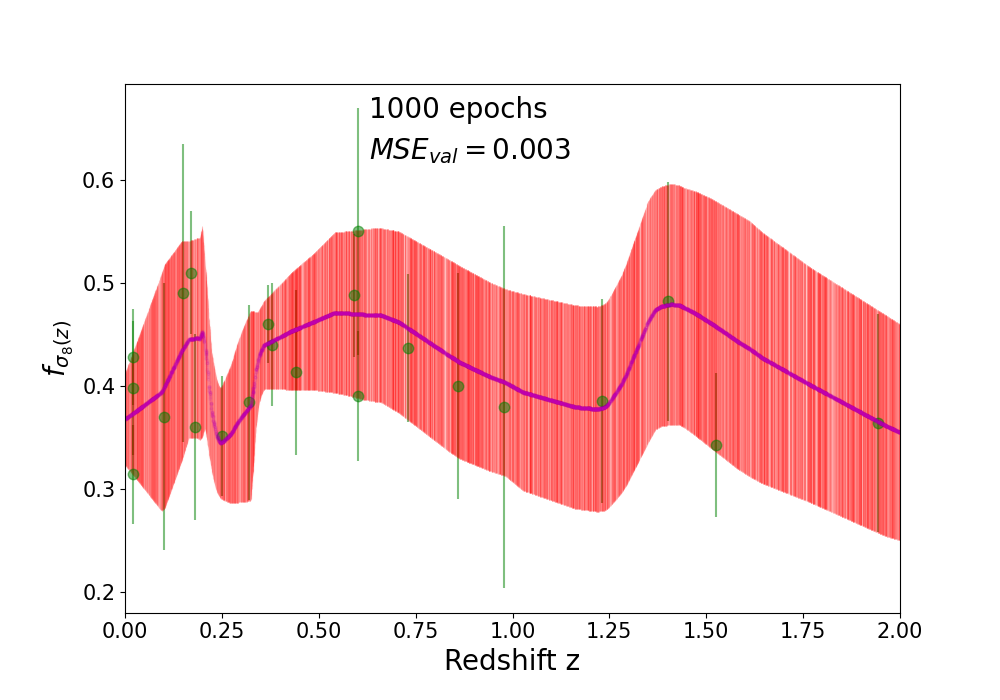}  
            }
    
    \caption{Reconstructions from Artificial Neural Networks trained with different numbers of epochs. The effect of the epochs can be noticed in training with the CC dataset (top) and with the $f{\sigma_8}(z)$ measurements (bottom). The first case  ($20$ epochs) shows underfitting while considering $1000$ epochs shows overfitting. In the $f{\sigma_8}$ dataset, the cases for $500$ and $1000$ epochs present overfitting. In both cases, we choose $100$ epochs due to the lowest value of MSE in the validation set. Green points display real data points with error bars, and in purple \changes{the neural reconstructions} along with red error bars.}
    \label{fig:epochs}
    \end{figure*}
    This appendix describes some aspects we consider in training  our feedforward neural networks. Although the goal of the neural network training is to minimize the loss function, {also} the following relationship should be taken into account:
        \begin{equation}
        	{\rm MSE} = {\rm bias}^2 + {\rm variance},
        \end{equation}
    where the bias measures how far away the neural network predictions are from the actual value, while the variance refers to how much the prediction varies at nearby points. As the ANN model gets more complex, the bias can decrease while the variance can increase. This is called the bias-variance dilemma \cite{geman1992neural}. A model with high variance will be overfitted, while a model with high bias will be insufficient to learn the complexity of the data (underfitting).
    
    In both cases, the model generated by the neural network would have inaccurate predictions. One way to avoid this problem is by monitoring the behavior of the loss function throughout the training epochs, both in the training set and in the validation set, both of them must have similar behavior. For example, in Figure \ref{fig:epochs}, we can see the effect of the number of epochs in two of our neural networks used in this work.
    A common practice to prevent an incorrect fitting in the ANN model is increasing the training set’s size. Otherwise, it is worth it to carefully calibrate the ANN models' hyperparameters to achieve acceptable results. 
    There are several approaches to tune the hyperparameters \cite{larochelle2007empirical, hutter2009paramils, bardenet2013collaborative, zhang2019deep}. Because our ANNs have relatively simple architectures (between two and five hidden layers and just a few thousand neurons), we use a common empirical strategy based on a grid of hyperparameters \cite{larochelle2007empirical} {where several combinations of hyperparameters are evaluated to choose the best of them}.
    In general, for the three types of cosmological observations (CC, $f{\sigma_8}$ and SNeIa), we have followed the next steps to find out a suitable neural network model for the corresponding data:
    \begin{itemize}
        \item We train several neural network configurations to gain insights into the complexity of their architecture to model the data. According to the loss function results, we choose a number of layers.
        \item Several values are suggested for each hyperparameter of the neural network. Based on the intuition achieved in the first step, a grid is formed that must be traversed to find the combination that provides the minimum value of the loss function. Among the hyperparameters are the batch size, the number of nodes per layer, and, in some cases, the dropout. 
        \item The best FNNN architectures found for each case are shown in Figure \ref{fig:arch}. The first two correspond to the CC and $f{\sigma_8}$ datasets, respectively, for which $320$ combinations were tested up to three hidden layers: number of nodes in $\{50, 100, 150, 200\}$ and the batch size in $\{1,4,8,16,32\}$. We found that for the compressed JLA dataset, a one-layer neural network works best, so we refined the third architecture among 20 combinations, varying the number of nodes in $\{30, 50, 100, 150, 200\}$ and the batch size in $\{1, 2, 4, 8\}$.
        \item We train the neural network with the combination of hyperparameters chosen in the previous step with a correct number of epochs. We verify the behavior of the loss function in the training and validation sets to check that our model is neither underfitted nor overfitted.
        %
        \item Once the neural network is trained, we can generate \changes{model-independent reconstructions with several predictions} and compare it with the original data.
        %
        %
        \item We compare the parameter estimation \changes{using data points from reconstructions} with the original datasets to verify they are statistically consistent; \changes{if they are not, the neural networks must be retrained or other architecture to be used.} For Bayesian inference, we use the \texttt{SimpleMC}\footnote{www.github.com/ja-vazquez/SimpleMC} package, initially released at \cite{aubourg2015}, along with a modified version of the \texttt{dynesty} nested sampling library \cite{speagle2020dynesty}, which allows to do the parameter estimation.
    \end{itemize}

    On the other hand, through the analysis of the JLA SNeIa compilation, we also use an FFNN to learn the behavior of the data from measurements of the distance modulus in a similar fashion we did for the CC and $f{\sigma_8}$. However, to handle the full covariance matrix, we use a VAE as described in the Appendix \ref{sec:appendix_vae_train}; using this type of neural network allows us to map the distribution of the distance modulus data to the distribution of the coded representation of the autoencoder to generate new covariance matrices. 
    One restriction of this method to bear in mind is that the new matrix must have the same dimension as the original one. However, we can generate any matrix given a combination of new redshifts, provided that this set has the same length as the original measurements. 

    In addition to the above procedure, we slightly modify the FFNNs to implement MC-DO. In this way, we add dropout between the layers of the FFNNs and run the MC-DO several times to obtain average values and uncertainties for each prediction (as described in Section \ref{sec:ann}). We combine our FFNN designs with the implementation of MC-DO layers from \texttt{astronn} \cite{leung2019deep} and compare the results of this method with the previous ANNs implementations.
    Because dropout is a regularization technique, the number of epochs is irrelevant for a large enough set.
    The error predictions and the uncertainties are independent; therefore, the total standard deviation is:
    \begin{equation}
        \sigma = \sqrt{er_{p}+\sum_i u_i^2 },
        \label{eq:sigma}
    \end{equation}
    \noindent
    where $u_i$ is the epistemic uncertainty involved with the FFNN used and  $er_{p}$ is the error prediction. 

    Besides the intrinsic error associated with the datasets, we consider an uncertainty related to the FFNN by adding a Monte Carlo dropout between each layer of the chosen FNNN architecture. Among several tests to dropout values between $[0, 0.1, 0.2, 0.3, 0.4, 0.5]$, we evaluated the loss function of the neural networks trained with these dropout values and found that the lower value of the loss function in the test and validation sets is with a dropout of $0.3$ along $1000$ epochs to the cosmic chronometers data, $0.1$ along $2000$ epochs to $f_{\sigma 8}$ data, and $0.01$ to the SNeIa data with $1000$ epochs. After training the FFNNs with MC-DO, we made $100$ executions of MC-DO for each prediction.

\section{Variational Autoencoder for non-diagonal covariance matrix}
\label{sec:appendix_vae_train}

\begin{figure*}[h!]
        \captionsetup{justification=raggedright,singlelinecheck=false,
        font=footnotesize}
        \centering
        \makebox[12cm][c]{
        \includegraphics[trim= 50mm 20mm 55mm 72mm, clip, width=5.5cm, height=4.0cm]{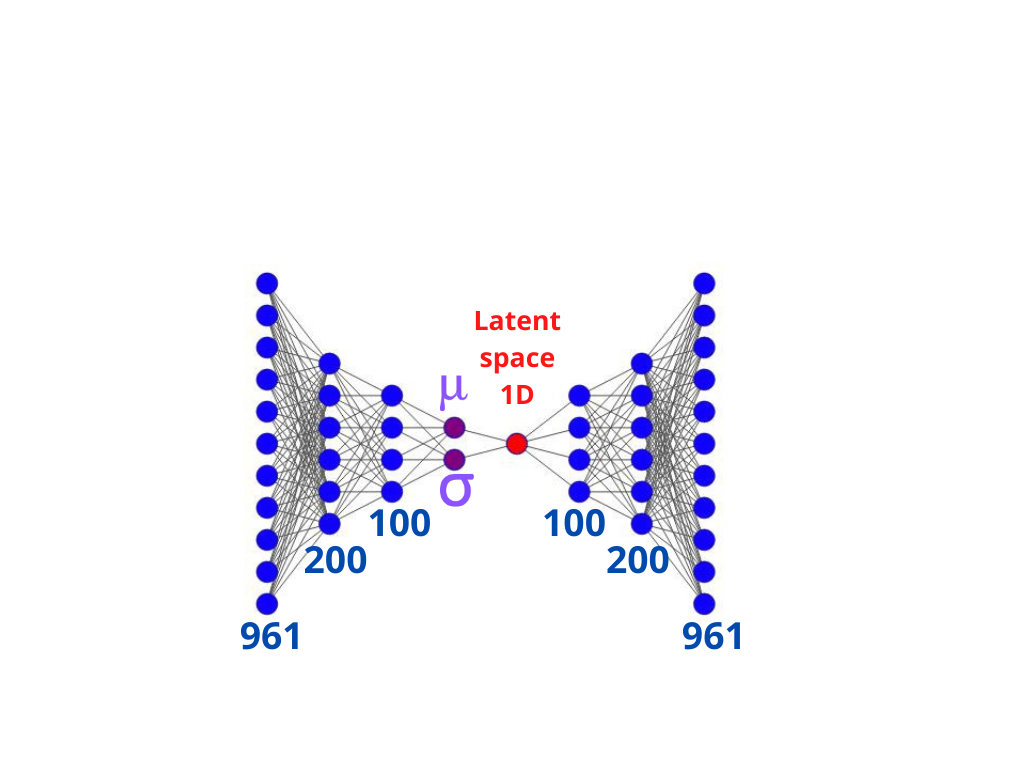}
        \includegraphics[trim= 0mm 0mm 0mm 0mm, clip, width=5.5cm, height=4.5cm]{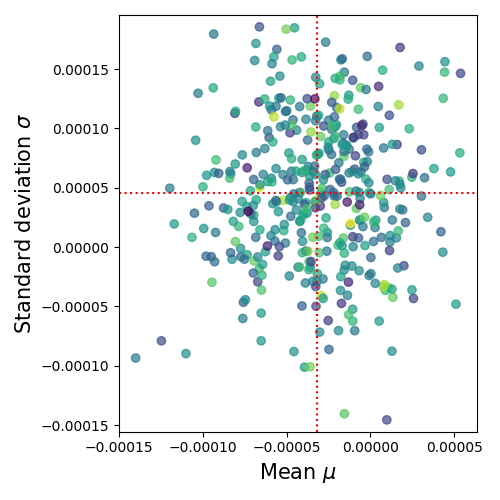}
        }
        \makebox[12cm][c]{
        \includegraphics[trim= 10mm 0mm 0mm 10mm, clip, width=5.5cm, height=4.5cm]{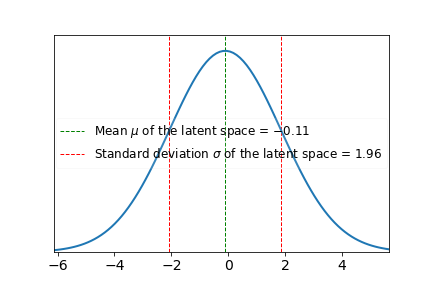}
        \includegraphics[trim= 0mm 0mm 10mm 10mm, clip, width=5.5cm, height=4.5cm]{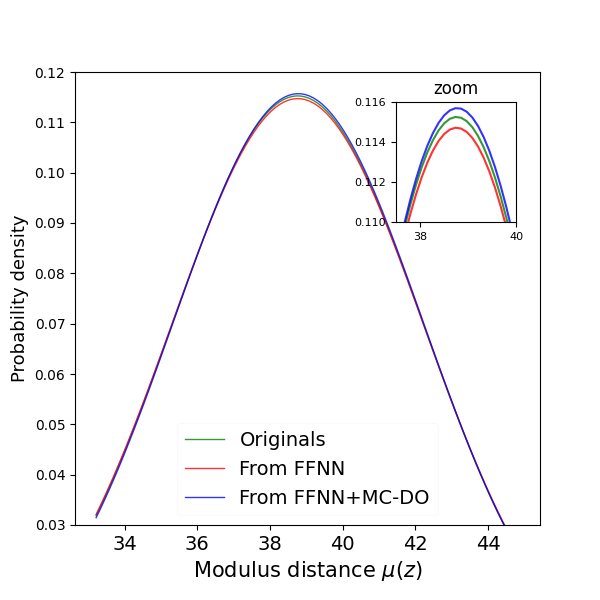}
        }
    \caption{\footnotesize{VAE results. \textit{Top left}: VAE architecture. \textit{Top right}: Samples of the mean and variance layers. \textit{Lower left}: Sampled distribution of the latent space. \textit{Lower right}: Comparison between the distributions for the modulus distances from different sources mapped into the latent space to generate a new covariance matrix with the VAE decoder.}}
    \label{fig:vae_cov}
\end{figure*}
    In this appendix, we explain a new method used to generate synthetic covariance matrices from the original matrix of the JLA SNeIa binned version. When the VAE is trained, it samples the probability distribution of the latent space, and because this covariance matrix is associated with the distance modulus measurements, therefore, we map its distribution to the latent space distribution.

    To have a dataset to train our VAE, we generated thousands of matrices by adding Gaussian noise of the same order of magnitude for each entry of the original covariance matrix. We assume that predictions at new redshifts within the current range of redshifts could have a similar covariance matrix, this assumption may be useful to test the method in a first approximation. {We designed the VAE architecture, shown in the first panel of Figure \ref{fig:vae_cov}}, to generate synthetic covariance matrices from a single point in the latent space. This VAE was trained on a dataset created from the systematic error covariance matrix of the JLA binned version. $\mu$ and $\sigma$ represent two layers connected to the last layer of the encoder and the latent space; in this case, both layers have a single neuron (the same dimension as the latent space). We use a batch size of 32 and the hyperbolic tangent as the activation function.

    Since we are interested on mapping the distribution of the distance modulus to the latent space, we design the VAE with a 1-dimensional latent space, so its mean $\mu$ and variance $\sigma$ are also 1-dimensional. After training the VAE, as in other generative neural networks, we can use the decoder part to generate new covariance matrices that traverse the latent space. Once the VAE is trained, we can explore the mean, variance, and latent space layers, as seen in the second and third panels of Figure \ref{fig:vae_cov}. To generate covariance matrices from the predictions of the modular distances coming from the FFNNs, using their means and standard deviations, we have assigned them a Gaussian distribution (fourth panel in Figure \ref{fig:vae_cov}). We have related the original measurements to the most likely region of the latent space. The deviations from the original measurements can be linearly mapped to the latent space to generate a new covariance matrix, as shown in Figure \ref{fig:covmatrix}.
    
    In our case, the VAE only learns to generate 31x31 covariance matrices, so we can only generate 31 predicted SNeIa sets. We use a variational autoencoder because the latent space has a probability distribution sampled during training. We map this distribution and the distance modulus using the decoder map. We generate a new covariance matrix for a different modulus of distance values; it is an experimental procedure, but it seems to work.

\end{document}